\documentclass[11pt]{article}

\usepackage[final]{acl}

\usepackage{times}
\usepackage{latexsym}
\usepackage{tcolorbox}
\tcbuselibrary{breakable,skins}

\usepackage[T1]{fontenc}

\usepackage[utf8]{inputenc}

\usepackage{microtype}

\usepackage{inconsolata}

\usepackage{graphicx}

\usepackage{bm}
\usepackage{booktabs}
\usepackage{multirow}
\usepackage{amsmath}
\usepackage{colortbl}
\usepackage{xcolor}
\usepackage{algorithm}
\usepackage{amssymb}
\usepackage{algpseudocode}
\usepackage{subcaption}
\usepackage{enumitem}       
\usepackage{titletoc}  
\usepackage{pifont}    
\newcommand{\hlc}[2][yellow]{{\setlength{\fboxsep}{0.5pt}\colorbox{#1!70!white}{#2}}}
\newcommand{\cmark}{\ding{51}}
\definecolor{relcolor}{HTML}{E6F5EA}  

\hypersetup{hidelinks}

\title{HeadRank: Decoding-Free Passage Reranking via Preference-Aligned Attention Heads}

\author{
 \textbf{Juyuan Wang\textsuperscript{1,$*$}},
 \textbf{Chenxing Wang\textsuperscript{1,$*$,$\S$}},
  \textbf{Yuchen Fang\textsuperscript{1}},
 \textbf{Huiyun Hu\textsuperscript{1}},
 \textbf{Junwu Du\textsuperscript{1}},
 \\
  \textbf{Aolin Li\textsuperscript{1}},
  \textbf{Shunlin Rong\textsuperscript{1}},
  \textbf{Haijun Wu\textsuperscript{1}},
 \textbf{Jin Xu\textsuperscript{2}},
 \textbf{Ligang Liu\textsuperscript{1}},
 \textbf{Dongliang Liao\textsuperscript{2,$\dagger$}}
\\
\\
 \textsuperscript{1}Weixin Group, Tencent, China\\
 \textsuperscript{2}South China University of Technology, Guangzhou, China
\\
 \small{
   \textbf{$*$} Equal contribution \quad
   \textbf{$\S$} Project lead \quad
   \textbf{$\dagger$ Correspondence:} \href{mailto:liaodl@scut.edu.cn}{liaodl@scut.edu.cn}
 }
}

\begin{document}
\maketitle
\begin{abstract}Decoding-free reranking methods that read relevance signals directly from LLM attention weights offer significant latency advantages over autoregressive approaches, yet suffer from attention score homogenization: middle-context documents receive near-identical scores, destroying the fine-grained distinctions required for ranking.
We propose HeadRank, a framework that lifts preference optimization from discrete token space into the continuous attention domain through entropy-regularized head selection, hard adjacent-level preference pairs, and a distribution regularizer that jointly sharpen discriminability in the homogenized middle zone.
Depth truncation at the deepest selected layer further reduces inference to $\mathcal{O}(1)$ forward passes.
Across 14 benchmarks on three Qwen3 scales (0.6B--4B) using only 211 training queries, HeadRank achieves the highest average NDCG@10 at every scale, outperforming both generative and decoding-free baselines on the majority of benchmarks with 100\% formatting success.
At 4B, 57.4\% of relevant middle-zone documents reach the top quartile versus 14.2\% for irrelevant ones---a 43-percentage-point selectivity gap that demonstrates the effectiveness of attention-space preference alignment for listwise reranking.
\end{abstract}
\section{Introduction}
\label{sec:introduction}

Reranking with Large Language Models (LLMs) has become a dominant paradigm in information retrieval~\cite{zhu2023large}. Generative approaches---listwise~\cite{sun2023chatgpt}, pointwise~\cite{nogueira2020document}, pairwise~\cite{qin2024large}, and setwise~\cite{zhuang2024setwise}---require at least one decoding step per candidate, incurring latency that scales with list length and risking malformed outputs that silently corrupt rankings.

\begin{figure}[t]
  \centering
  \includegraphics[width=\columnwidth]{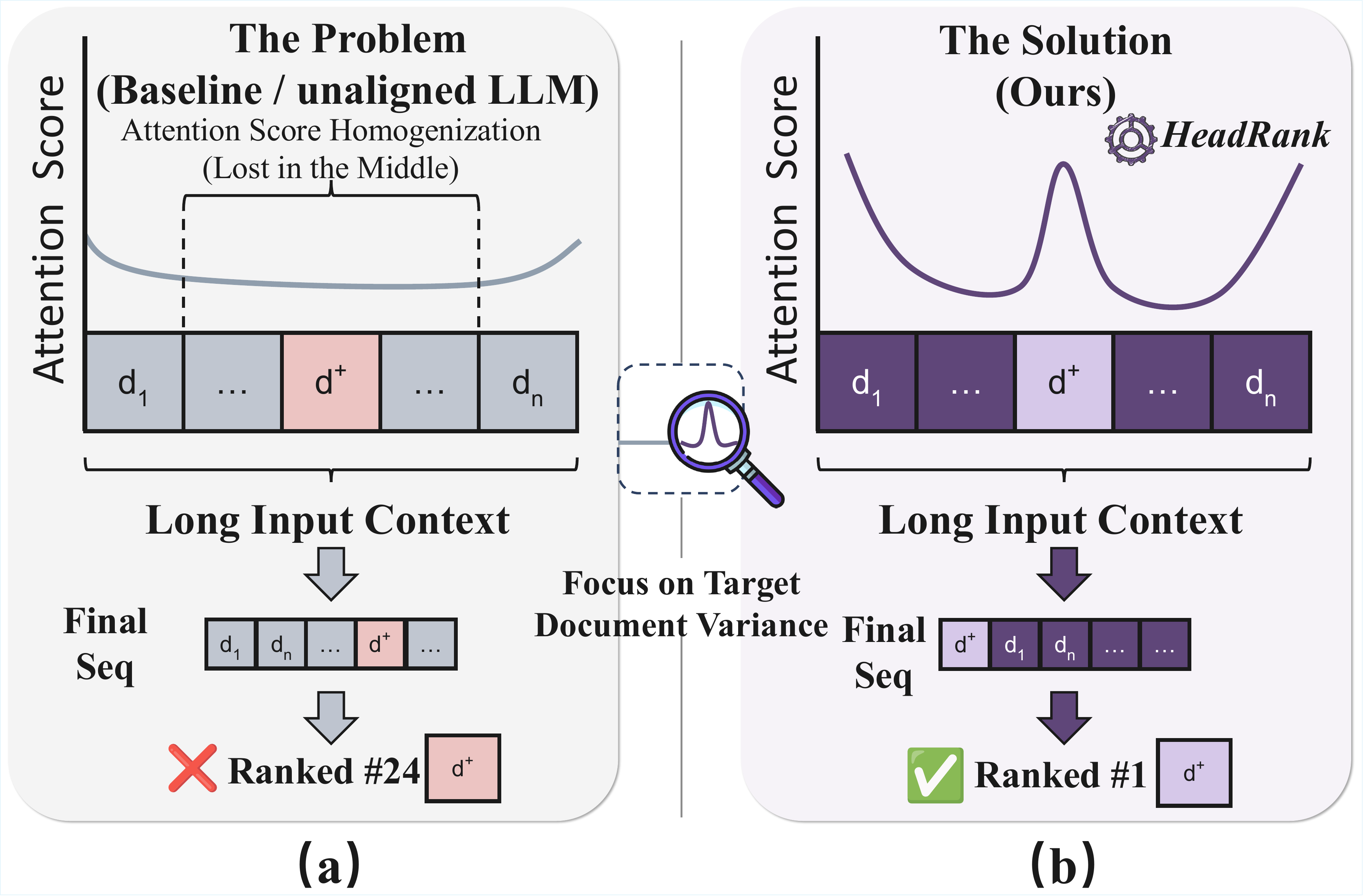}
  \caption{%
  \textbf{Comparison of reranking paradigms.}
  }
  \label{fig:overview}
  \vspace{-1em}
\end{figure}

A recent line of work sidesteps generation entirely by reading relevance signals from attention weights during the prefill pass~\cite{icr, core, mvp2025}---no decoding, no formatting risk. However, we find that these methods hit a wall as the candidate list grows due to what we call \textit{attention score homogenization}: attention weights for middle-context documents become nearly uniform, destroying the fine-grained distinctions that ranking requires. Current head-selection heuristics~\cite{core, qrr} exacerbate this by latching onto noisy, high-entropy heads never trained for ranking.

We resolve this with \textbf{HeadRank}, which combines three complementary ideas: (i)~entropy-regularized head selection that isolates stable, low-entropy retrieval heads; (ii)~Adjacent-Level Preference Sampling (ALPS) that constructs hard contrastive pairs from adjacent relevance grades; and (iii)~an attention-space DPO objective with a distribution regularizer that explicitly fights score flatlining. At inference time, the forward pass is truncated at the deepest selected layer $l_{\max}$, yielding $\mathcal{O}(1)$-pass complexity.
Figure~\ref{fig:overview} contrasts the failure mode with our solution: baseline methods produce flat attention curves that bury relevant documents in the middle zone, whereas HeadRank restores discriminative peaks.

Our main contributions are twofold:
\textbf{(1)} We identify \textit{attention score homogenization} as the mechanistic bottleneck of decoding-free reranking and propose HeadRank, which lifts DPO into continuous attention space with entropy-regularized head discovery and Adjacent-Level Preference Sampling (ALPS), enabling preference alignment without autoregressive decoding. A depth-truncation strategy further reduces inference cost to $l_{\max}/L$ of a full forward pass.
\textbf{(2)} Experiments on 14 benchmarks show that HeadRank trained on only 211 queries achieves the best average performance across both generative and decoding-free baselines at all three model scales (0.6B--4B), with 100\% formatting success rate and substantial latency savings.

\section{Related Work}
\label{sec:related_work}

\textbf{LLM-Based Re-ranking.}
LLM reranking methods are broadly categorized into generative and attention-based paradigms.
Generative approaches prompt LLMs to produce ranked outputs through autoregressive decoding: RankGPT~\cite{sun2023chatgpt} outputs listwise permutations, while pointwise~\cite{nogueira2020document}, pairwise~\cite{qin2024large}, and setwise~\cite{zhuang2024setwise} variants reduce output length but still require at least one decoding step per candidate.
Recent efforts apply reinforcement learning to generative rerankers~\cite{zhuang2025rankr1, zhang2025rearank}, yet these methods remain in token space and inherit decoding overhead and formatting risks.
In contrast, attention-based reranking sidesteps generation entirely.
ICR~\cite{icr} first demonstrated that aggregated attention scores from a single prefill pass can serve as relevance signals, eliminating decoding latency.
Subsequent work identifies specialized ``retrieval heads'' responsible for information recall~\cite{niah, yin2025attention}: CoRe~\cite{core} discovers such heads via a contrastive metric, QR-R~\cite{qrr} uses causal interventions, and NIAH~\cite{niah} leverages needle-in-a-haystack probing.
However, all existing attention-based methods select heads through post-hoc heuristics without explicit preference optimization, leaving them vulnerable to score homogenization when the candidate list grows.

\textbf{Attention Patterns and Positional Bias in LLMs.}
The premise that attention weights encode relevance is empirically supported~\cite{serrano2019attention, wiegreffe2019attention}, and mechanistic interpretability has revealed functional specialization among attention heads---with only a sparse subset being critical for specific tasks~\cite{niah, yin2025attention}.
A closely related challenge is the ``Lost in the Middle'' phenomenon~\cite{liu2024lost, peysakhovich2023attention}, where LLMs systematically under-attend to middle-context information.
This behavior is linked to ``attention sinks'' that divert probability mass to structurally prominent tokens~\cite{xiao2023efficient}.
Permutation ensembling~\cite{tang2023found} mitigates positional bias but multiplies latency proportionally.
Our work provides a mechanistic explanation for this phenomenon at the level of individual attention heads---what we term \textit{attention score homogenization}---and directly resolves it through preference alignment rather than inference-time ensembling.

\textbf{Preference Optimization for Ranking.}
Direct Preference Optimization (DPO)~\cite{rafailov2024direct} aligns LLMs via token-level generation probabilities without explicit reward models, and has been adapted for generative reranking.
However, applying DPO directly to reranking requires autoregressive decoding, defeating the latency benefits of attention-based approaches.
HeadRank projects preference alignment from discrete token space into continuous attention space, enabling DPO-style optimization over attention head activations without any generation step.

\section{Methodology}
\label{sec:methodology}

\definecolor{baselinecolor}{HTML}{F5F5F5}      
\definecolor{ourscolor}{HTML}{EFE4F7}          
\begin{table*}[t]
\centering
\caption{NDCG@10 across in-domain and out-of-domain benchmarks with BM25 top-40 retrieval across Qwen3 model sizes. \textbf{SR (\%)} denotes the formatting Success Rate of the generated outputs.}
\label{tab:ndcg10_top40}
\setlength{\tabcolsep}{4pt}
\resizebox{\textwidth}{!}{%
\begin{tabular}{@{} ll *{11}{c} | cc @{}}
\toprule
& & \multicolumn{2}{c}{\textbf{In-Domain}} & \multicolumn{9}{c}{\textbf{Out-of-Domain}} & & \\
\cmidrule(lr){3-4} \cmidrule(lr){5-13}
\textbf{Model} & \textbf{Method} & \textbf{DL19} & \textbf{DL20} & \textbf{COVID} & \textbf{ArguAna} & \textbf{News} & \textbf{FiQA} & \textbf{SciDocs} & \textbf{NFC} & \textbf{NQ} & \textbf{DBPedia} & \textbf{FEVER} & \textbf{Avg.} & \textbf{SR (\%)} \\
\midrule
\textit{Baseline}
 & \cellcolor{baselinecolor}BM25 (top40) & \cellcolor{baselinecolor}50.58 & \cellcolor{baselinecolor}47.96 & \cellcolor{baselinecolor}59.47 & \cellcolor{baselinecolor}29.99 & \cellcolor{baselinecolor}39.52 & \cellcolor{baselinecolor}23.61 & \cellcolor{baselinecolor}14.90 & \cellcolor{baselinecolor}33.75 & \cellcolor{baselinecolor}30.55 & \cellcolor{baselinecolor}31.80 & \cellcolor{baselinecolor}65.13 & \cellcolor{baselinecolor}38.84 & \cellcolor{baselinecolor}100.0 \\
\midrule
\multirow{6}{*}{\textit{Qwen3-0.6B}}
 & RankGPT & 49.17 & \underline{48.54} & 60.28 & 26.44 & 37.81 & 23.22 & 14.60 & 33.48 & 30.89 & 32.18 & 60.85 & 37.95 & 76.6 \\
 & ICR & 50.09 & 41.75 & 63.28 & 32.43 & 41.13 & 22.70 & 13.75 & 34.06 & 33.62 & 32.57 & 65.59 & 39.18 & 100.0 \\
 & CoRe & 50.46 & 44.24 & 65.53 & 33.61 & \underline{42.55} & \underline{25.66} & \underline{15.48} & 33.22 & \underline{36.61} & \underline{34.20} & 71.00 & \underline{41.14} & 100.0 \\
 & QR-R & \underline{51.96} & 43.32 & \underline{67.84} & \underline{34.39} & 40.04 & 24.42 & 13.93 & \underline{34.28} & 35.45 & 33.22 & 66.18 & 40.46 & 100.0 \\
 & NIAH & 46.55 & 38.87 & 59.76 & 33.87 & 39.34 & 23.53 & 14.58 & 32.40 & 33.83 & 31.82 & \underline{71.65} & 38.75 & 100.0 \\
 & \cellcolor{ourscolor}HeadRank & \cellcolor{ourscolor}\textbf{59.93} & \cellcolor{ourscolor}\textbf{51.71} & \cellcolor{ourscolor}\textbf{68.71} & \cellcolor{ourscolor}\textbf{34.62} & \cellcolor{ourscolor}\textbf{44.02} & \cellcolor{ourscolor}\textbf{25.99} & \cellcolor{ourscolor}\textbf{16.36} & \cellcolor{ourscolor}\textbf{35.09} & \cellcolor{ourscolor}\textbf{43.25} & \cellcolor{ourscolor}\textbf{37.02} & \cellcolor{ourscolor}\textbf{73.47} & \cellcolor{ourscolor}\textbf{44.56} & \cellcolor{ourscolor}\textbf{100.0} \\
\midrule
\multirow{6}{*}{\textit{Qwen3-1.7B}}
 & RankGPT & \underline{56.32} & \underline{52.77} & 68.15 & 27.47 & \underline{46.30} & 25.95 & 16.46 & 28.82 & 39.50 & \underline{36.44} & 54.69 & 41.17 & 97.1 \\
 & ICR & 52.11 & 46.14 & 65.42 & 30.63 & 43.47 & 27.17 & 15.59 & 34.65 & 37.09 & 33.24 & 69.39 & 41.35 & 100.0 \\
 & CoRe & 51.26 & 45.81 & \underline{68.67} & \underline{34.59} & 42.71 & \textbf{28.49} & \underline{16.57} & \underline{34.91} & \underline{40.00} & 34.39 & 71.02 & \underline{42.58} & 100.0 \\
 & QR-R & 52.36 & 46.56 & 67.77 & 33.52 & 43.89 & 27.61 & 16.00 & 34.70 & \underline{40.00} & 33.36 & 65.45 & 41.93 & 100.0 \\
 & NIAH & 48.98 & 43.25 & 63.39 & 33.00 & 41.22 & 26.29 & 15.44 & 33.42 & 35.98 & 32.15 & \textbf{72.59} & 40.52 & 100.0 \\
 & \cellcolor{ourscolor}HeadRank & \cellcolor{ourscolor}\textbf{61.32} & \cellcolor{ourscolor}\textbf{57.85} & \cellcolor{ourscolor}\textbf{70.81} & \cellcolor{ourscolor}\textbf{36.59} & \cellcolor{ourscolor}\textbf{46.40} & \cellcolor{ourscolor}\underline{28.11} & \cellcolor{ourscolor}\textbf{16.96} & \cellcolor{ourscolor}\textbf{36.23} & \cellcolor{ourscolor}\textbf{43.32} & \cellcolor{ourscolor}\textbf{37.06} & \cellcolor{ourscolor}\underline{71.86} & \cellcolor{ourscolor}\textbf{46.05} & \cellcolor{ourscolor}\textbf{100.0} \\
\midrule
\multirow{6}{*}{\textit{Qwen3-4B}}
 & RankGPT & \underline{63.58} & \underline{59.52} & 71.12 & 27.72 & \textbf{49.32} & 27.23 & \underline{16.68} & 30.24 & \underline{44.86} & \textbf{41.34} & 62.18 & \underline{44.89} & 95.8 \\
 & ICR & 54.38 & 48.17 & 66.63 & 23.68 & 42.69 & 24.53 & 14.47 & 34.61 & 37.67 & 32.81 & 61.06 & 40.06 & 100.0 \\
 & CoRe & 56.67 & 50.89 & 70.08 & \underline{32.93} & 43.32 & \textbf{29.88} & 16.52 & 35.13 & 42.80 & 35.84 & \textbf{73.39} & 44.31 & 100.0 \\
 & QR-R & 57.45 & 53.84 & \underline{73.04} & 32.35 & 44.46 & 28.18 & 15.98 & \underline{35.68} & 44.30 & 36.27 & 69.82 & 44.67 & 100.0 \\
 & NIAH & 48.13 & 40.94 & 63.64 & 24.91 & 38.30 & 25.00 & 14.69 & 32.26 & 34.40 & 30.82 & \underline{70.83} & 38.54 & 100.0 \\
 & \cellcolor{ourscolor}HeadRank & \cellcolor{ourscolor}\textbf{64.33} & \cellcolor{ourscolor}\textbf{59.89} & \cellcolor{ourscolor}\textbf{73.94} & \cellcolor{ourscolor}\textbf{35.30} & \cellcolor{ourscolor}\underline{46.27} & \cellcolor{ourscolor}\underline{28.73} & \cellcolor{ourscolor}\textbf{16.89} & \cellcolor{ourscolor}\textbf{36.44} & \cellcolor{ourscolor}\textbf{45.36} & \cellcolor{ourscolor}\underline{37.24} & \cellcolor{ourscolor}69.69 & \cellcolor{ourscolor}\textbf{46.73} & \cellcolor{ourscolor}\textbf{100.0} \\
\bottomrule
\end{tabular}%
}
\end{table*}

\subsection{Problem Formulation and Preliminaries}
\label{subsec:preliminaries}

Document reranking aims to compute a relevance score $s(q, d_i)$ for each document in a candidate set $\mathcal{D} = \{d_1, d_2, \dots, d_N\}$, which is typically retrieved by an initial ranker such as BM25, to produce a sorted permutation. Modern LLMs approach this task through stacked Transformer layers, whose internal attention distributions provide a natural relevance signal. 

Generative reranking methods inherently rely on computationally expensive autoregressive decoding. In contrast, attention-based approaches utilize internal activations during the prefill phase. Following the paradigm of In-Context Re-ranking (ICR), the input sequence is formatted by placing documents before the query, denoted as $X = [\text{Instruction} \oplus d_1 \oplus \dots \oplus d_N \oplus q]$, to satisfy the causal attention mask requirements of decoder-only LLMs. This arrangement ensures that query tokens can attend to preceding document tokens. 

For a specific attention head $h$ in layer $l$, let $A_{i,j}^{(l,h)}$ denote the attention weight from the $i$-th token to the $j$-th token, quantifying the pairwise information flow. Rather than relying on standard decoding, we directly extract these internal distributions. The relevance score $\alpha_d^{(l,h)}$ for a specific head $(l,h)$ is defined as the aggregated attention mass from query tokens to document tokens:
\begin{equation}
    \alpha_d^{(l,h)} = \frac{1}{|\mathcal{I}_q|} \sum_{i \in \mathcal{I}_q} \sum_{j \in \mathcal{I}_d} A_{i,j}^{(l,h)}
    \label{eq:attention_score}
\end{equation}
where $\mathcal{I}_q$ and $\mathcal{I}_d$ denote the token index sets for the query and the document, respectively.

\subsection{Entropy-Regularized Core Head Selection}
\label{subsec:core_head_selection}

Identifying the right attention heads is critical: contrastive metrics alone can select heads that are nominally discriminative but exhibit chaotic, high-entropy distributions, introducing noise during alignment. Concurrent work on retrieval-head contrasting for hallucination mitigation underscores the importance of principled head identification. We therefore propose an entropy-regularized discovery mechanism.

Given a relevant document $d^+$ and a set of irrelevant documents $D^-$, we follow CoRe~\cite{core} and first compute a base discriminative score $S_{\text{disc}}^{(l,h)}$ for head $h$ in layer $l$. This score employs a temperature-scaled softmax function to measure the effectiveness of the head in distinguishing the positive document from the negative documents:
\begin{equation}
    S_{\text{disc}}^{(l,h)} = \frac{e^{\alpha_{d^+}^{(l,h)} / \tau}}{ \sum_{d \in \{d^+\} \cup D^-} e^{\alpha_d^{(l,h)} / \tau} }
\end{equation}
To penalize heads with dispersed attention, we define an entropy gating factor. Let $A_t^{(l,h)} = \frac{1}{|\mathcal{I}_q|}\sum_{i \in \mathcal{I}_q} A_{i,t}^{(l,h)}$ denote the query-averaged attention weight to the $t$-th token, obtained by marginalizing $A_{i,j}^{(l,h)}$ over query positions. The Shannon entropy of this distribution over the sequence length $L_{\text{seq}}$ is denoted as $\mathcal{H}^{(l,h)} = -\sum_{t=1}^{L_{\text{seq}}} A_t^{(l,h)} \log A_t^{(l,h)}$. This value is normalized by the maximum possible entropy $\log L_{\text{seq}}$ to define the gating factor $G_{\text{ent}}^{(l,h)}$:
\begin{equation}
    G_{\text{ent}}^{(l,h)} = 1 - \lambda \frac{\mathcal{H}^{(l,h)}}{\log L_{\text{seq}}}
\end{equation}
where $\lambda$ represents a hyperparameter that scales the penalty.

The final selection criterion $\Phi(l,h)$ combines both components to favor heads with a sharp focus while downplaying those with high-entropy distributions:
\begin{equation}
    \Phi(l,h) = S_{\text{disc}}^{(l,h)} \cdot G_{\text{ent}}^{(l,h)}
    \label{eq:entropy_selection}
\end{equation}
The top $K$ heads are selected based on $\Phi(l,h)$, and the maximum layer index among these heads determines the depth for inference termination, denoted as $l_{\max}$.

\begin{table*}[t]
\centering
\caption{Recall@2 and Recall@5 on knowledge-intensive multi-hop reasoning benchmarks (HotpotQA, 2WikiMultihopQA, MuSiQue) with ColBERTv2 top-20 retrieval. \textbf{SR} denotes the formatting Success Rate of generative methods. Reference rows (\textit{Mistral 7B}, \textit{Llama-3.1 8B}, \textit{GPT-3.5 Turbo}) run RankGPT on their respective models.}
\label{tab:multihop}
\renewcommand{\arraystretch}{1} 
\setlength{\tabcolsep}{4pt}
\small 
\begin{tabular}{@{} ll *{6}{c} | *{2}{c} c @{}}
\toprule
& & \multicolumn{2}{c}{\textbf{HotpotQA}} & \multicolumn{2}{c}{\textbf{2Wiki}} & \multicolumn{2}{c|}{\textbf{MuSiQue}} & \multicolumn{2}{c}{\textbf{Avg.}} & \\
\textbf{Model} & \textbf{Method} & \textbf{R@2} & \textbf{R@5} & \textbf{R@2} & \textbf{R@5} & \textbf{R@2} & \textbf{R@5} & \textbf{R@2} & \textbf{R@5} & \textbf{SR} \\
\midrule
\rowcolor{baselinecolor}  & ColBERTv2 & 64.65 & 79.30 & 59.23 & 68.20 & 37.94 & 49.52 & 53.94 & 65.67 & — \\
\midrule
\multirow{3}{*}{\textit{Qwen3-0.6B}}
 & RankGPT & 52.85 & 62.90 & 39.50 & 46.70 & 30.42 & 37.49 & 40.92 & 49.03 & 80.1\% \\
 & ICR & 57.50 & 75.35 & 51.28 & 62.95 & 31.23 & 43.69 & 46.67 & 60.66 & 100\% \\
 & \cellcolor{ourscolor}HeadRank & \cellcolor{ourscolor}\textbf{67.50} & \cellcolor{ourscolor}\textbf{85.25} & \cellcolor{ourscolor}\textbf{60.98} & \cellcolor{ourscolor}\textbf{72.12} & \cellcolor{ourscolor}36.72 & \cellcolor{ourscolor}\textbf{51.02} & \cellcolor{ourscolor}\textbf{55.07} & \cellcolor{ourscolor}\textbf{69.46} & \cellcolor{ourscolor}100\% \\
\midrule
\multirow{3}{*}{\textit{Qwen3-1.7B}}
 & RankGPT & 57.25 & 73.20 & 46.45 & 59.15 & 35.02 & 47.83 & 46.24 & 60.06 & 93.7\% \\
 & ICR & 66.95 & 82.10 & 59.65 & 70.60 & \textbf{39.61} & 51.07 & 55.40 & 67.92 & 100\% \\
 & \cellcolor{ourscolor}HeadRank & \cellcolor{ourscolor}\textbf{69.35} & \cellcolor{ourscolor}\textbf{86.20} & \cellcolor{ourscolor}\textbf{62.65} & \cellcolor{ourscolor}\textbf{72.88} & \cellcolor{ourscolor}39.05 & \cellcolor{ourscolor}\textbf{54.47} & \cellcolor{ourscolor}\textbf{57.02} & \cellcolor{ourscolor}\textbf{71.18} & \cellcolor{ourscolor}100\% \\
\midrule
\multirow{3}{*}{\textit{Qwen3-4B}}
 & RankGPT & 71.60 & 84.25 & \textbf{65.18} & 73.28 & 41.25 & 52.99 & 59.34 & 70.17 & 98.9\% \\
 & ICR & 57.45 & 74.20 & 53.93 & 66.00 & 34.23 & 46.52 & 48.54 & 62.24 & 100\% \\
 & \cellcolor{ourscolor}HeadRank & \cellcolor{ourscolor}\textbf{72.65} & \cellcolor{ourscolor}\textbf{86.45} & \cellcolor{ourscolor}63.60 & \cellcolor{ourscolor}\textbf{73.75} & \cellcolor{ourscolor}\textbf{42.67} & \cellcolor{ourscolor}\textbf{55.10} & \cellcolor{ourscolor}\textbf{59.64} & \cellcolor{ourscolor}\textbf{71.77} & \cellcolor{ourscolor}100\% \\
\midrule
\rowcolor{baselinecolor} \textit{Mistral 7B} & RankGPT & 64.80 & 79.80 & 56.30 & 67.10 & 36.20 & 48.70 & 52.40 & 65.20 & 20.5\% \\
\rowcolor{baselinecolor} \textit{Llama-3.1 8B} & RankGPT & 69.90 & 84.80 & 60.80 & 72.00 & 39.30 & 51.40 & 56.60 & 69.40 & 98.2\% \\
\rowcolor{baselinecolor} \textit{GPT-3.5 Turbo} & RankGPT & 71.80 & 84.90 & 61.60 & 72.80 & 38.70 & 54.10 & 57.40 & 70.60 & 98.6\% \\
\bottomrule
\end{tabular}
\end{table*}

\subsection{Attention-based Direct Preference Optimization (HeadRank)}
\label{subsec:attn_headrank}

Standard DPO~\cite{rafailov2024direct} aligns token-level generation probabilities, so applying it directly to reranking requires autoregressive decoding and its attendant latency. We therefore project the alignment paradigm from discrete token space into continuous attention space.

Let $s_\theta^{(d)}$ denote the relevance score computed exclusively by aggregating the attention mass from the selected core heads. For a preference pair, let $s_\theta^+$ and $s_\theta^-$ be the scores for the chosen and rejected documents, respectively. To prevent the objective from degenerating into disjoint heuristics, we unify the margin constraints and anti-homogenization penalties into a single loss. Let $\Delta s_\theta = s_\theta^+ - s_\theta^-$ denote the predicted score margin, and $\bm{\Delta}_{\text{ref}} = [s_\theta^+ - s_{\text{ref}}^+, s_\theta^- - s_{\text{ref}}^-]^\top$ denote the reference divergence vector. The unified HeadRank objective is formulated as:
\begin{equation}
    \mathcal{L}_{\text{total}} = \mathcal{L}_{\text{align}}(\Delta s_\theta) + \mathcal{L}_{\text{prox}}(\bm{\Delta}_{\text{ref}}) + \Omega(\bm{s}_\theta)
\end{equation}
where $\mathcal{L}_{\text{prox}}(\bm{\Delta}_{\text{ref}}) = \frac{\beta}{2}\|\bm{\Delta}_{\text{ref}}\|_2^2$ is the proximal penalty, and the pairwise alignment objective $\mathcal{L}_{\text{align}}$ and the distribution regularizer $\Omega(\bm{s}_\theta)$ are formally defined as:
\begin{align}
    \resizebox{0.2\linewidth}{!}{$\mathcal{L}_{\text{align}}(\Delta s_\theta)$} &= \resizebox{0.6\linewidth}{!}{$-\log \sigma(\Delta s_\theta) + [m - \Delta s_\theta]_+ - \alpha \Delta s_\theta$} \\
    \Omega(\bm{s}_\theta) &= \gamma \mathcal{H}(\bm{p}) - \eta \operatorname{Var}(\bm{s}_{\text{mid}})
\end{align}

Here, $[x]_+ = \max(0, x)$ denotes the standard hinge operator with margin constraint $m$, preventing overconfidence. The linear term $-\alpha \Delta s_\theta$ complements the sigmoid: as $\Delta s_\theta$ grows, $-\log\sigma(\cdot)$ saturates and its gradient vanishes, whereas the linear term maintains a steady gradient that continues to encourage margin expansion; the proximal penalty $\mathcal{L}_{\text{prox}}$ bounds this push by anchoring scores near the reference model. To maintain the pre-trained retrieval capabilities and prevent representation collapse, we introduce a \textbf{Proximal Policy Retention} ($\mathcal{L}_{\text{prox}}$) term. Rather than the KL divergence used in standard DPO, we adopt an $L_2$ proximal penalty $\|\bm{\Delta}_{\text{ref}}\|_2^2$, which prevents the aligned policy from deviating too far from the frozen reference in the continuous attention domain.

$\Omega(\bm{s}_\theta)$ is a distribution regularizer that directly targets "Lost in the Middle" score homogenization. $\mathcal{H}(\bm{p})$ is the Shannon entropy of the document score distribution $\bm{p}$; minimizing it sharpens the overall score spread. Simultaneously, maximizing $\operatorname{Var}(\bm{s}_{\text{mid}})$---the variance of intermediate document scores\footnote{The ``middle zone'' comprises documents at BM25 ranks in the 25\textsuperscript{th}--75\textsuperscript{th} percentile (positions 11--30 for top-40, positions 6--15 for top-20).}---breaks the score-flattening effect among middle-ranked candidates, restoring fine-grained discriminability across the full context window.

\subsection{Early-Exit Inference via Depth Truncation}
\label{subsec:inference_pruning}

HeadRank is entirely \textit{decoding-free}: relevance scores are extracted from attention weights during the prefill phase, bypassing autoregressive generation entirely.
Since all $K$ core heads reside at or below layer $l_{\max}$, the forward pass safely exits at that depth, consuming only $l_{\max}/L$ of the full-model FLOPs.
Together, the decoding-free architecture and early termination yield $\mathcal{O}(1)$-pass inference with zero formatting failures.

\subsection{Data Construction and Iterative Alignment}
\label{subsec:data_and_iterative_alignment}

To bypass the heavy data reliance of generative models and synergize with the distribution regularizer $\Omega(\bm{s}_\theta)$, an efficient data construction pipeline is established using 211 queries from MS MARCO V2. The top-100 documents are retrieved via BM25 to form the context. To ensure pair quality, a margin-aware Adjacent-Level Preference Sampling (ALPS) strategy is applied. Motivated by the finding that fine-grained relevance labels improve zero-shot LLM rankers~\cite{zhuang2023beyond}, preference pairs exhibiting a relevance gap larger than one level (e.g., 3 versus 0 or 3 versus 1) are strictly discarded, restricting the construction exclusively to adjacent-level pairs (e.g., 3 versus 2). This formulation forces the attention heads to learn fine-grained, discriminative ranking features, directly complementing the intermediate score variance objective.

These same pairs are used for both core head selection and optimization. Because the attention topology shifts during training, an iterative refinement step is employed. After an initial training round, we re-run head selection on the updated weights to recalibrate the core set, keeping it aligned with the post-training attention topology.
The complete training and inference procedure is summarized in Algorithm~\ref{alg:attn_headrank} (Appendix~\ref{sec:algorithm_appendix}).

\section{Experiments}
\label{sec:experiments}

\subsection{Experimental Setup}
\label{subsec:experimental_setup}

\textbf{Datasets and Metrics.} We evaluate reranking effectiveness, out-of-domain generalization, and long-context reasoning across 14 benchmarks, organized into three groups. The first group comprises standard passage retrieval benchmarks, specifically TREC-DL19 and TREC-DL20, to assess in-domain alignment. The second group evaluates cross-domain transferability using seven diverse datasets from the BEIR benchmark: TREC-COVID~\cite{voorhees2021trec}, ArguAna, TREC-News, FiQA, SciDocs, NFCorpus (NFC), and DBPedia. The third group evaluates knowledge-intensive and multi-hop reasoning capabilities over long contexts using Natural Questions (NQ), FEVER, HotpotQA, 2WikiMultihopQA (2Wiki), and MuSiQue. For the passage retrieval and out-of-domain benchmarks, as well as NQ and FEVER, the top-40 candidate documents are initially retrieved using BM25 \cite{robertson2009probabilistic}. For multi-hop reasoning datasets, we follow the methodology of Gutiérrez et al. \cite{gutierrez2024} and employ ColBERTv2 \cite{santhanam2022colbertv2} to retrieve the top-20 candidate passages. For evaluation metrics, Normalized Discounted Cumulative Gain at rank 10 (NDCG@10) serves as the primary metric for the BM25-retrieved datasets, while Recall@2 and Recall@5 are measured for the ColBERTv2-retrieved multi-hop datasets. We also report Formatting Success Rate (SR) for generative methods---the fraction of queries yielding a correctly formatted ranking---to quantify output reliability.

\textbf{Baselines.} The proposed HeadRank framework is compared against five state-of-the-art reranking methods, encompassing both generative and decoding-free attention-based paradigms. For generative reranking, \textbf{RankGPT} \cite{sun2023chatgpt} is adopted as a representative baseline, outputting sorted permutations via autoregressive decoding (prompt template in Appendix~\ref{sec:prompt_appendix}). Supervised fine-tuned rerankers such as RankLLaMA~\cite{ma2024fine} and query-likelihood approaches~\cite{zhuang2023open} are orthogonal to our decoding-free paradigm and therefore excluded from the main comparison. For decoding-free attention-based methods, we compare against four closely related approaches. \textbf{ICR} \cite{icr} serves as the foundational baseline by uniformly aggregating global attention scores. \textbf{NIAH} \cite{niah} evaluates reranking using specific heads identified via the Needle-In-A-Haystack capability. \textbf{QR-R} \cite{qrr} leverages causal interventions to isolate query-dependent attention pathways. Finally, \textbf{CoRe} \cite{core} represents the current state-of-the-art, discovering retrieval heads based on a discrete contrastive metric without continuous alignment optimization.

\textbf{Implementation Details.}
We build HeadRank on three Qwen3 checkpoints~\cite{qwen3} (0.6B, 1.7B, and 4B parameters).
Training data consists of 211 queries from the MS~MARCO~V2 passage corpus; for each query, ALPS constructs adjacent-level
preference pairs from BM25 top-100 candidates, yielding ${\sim}91$K training pairs
for the 0.6B/1.7B models and ${\sim}200$K for the 4B model (the larger model's richer attention topology produces more head-level contrasts that pass the entropy gate, expanding the effective pair set).
We select $K{=}8$ core retrieval heads via entropy gating;
head distributions before and after alignment are reported in Appendix~\ref{sec:head_appendix}.
All models are trained for a single epoch.
Because several attention-based baselines (e.g., ICR) rely on calibration~\cite{zhao2021calibrate} to remove positional bias, we apply the same calibration protocol uniformly across all methods---including HeadRank---at both training and inference time, ensuring a fair comparison.
At inference, all attention-based methods process the
full BM25 top-40 list in a single forward pass with batch size 16, avoiding
the latency and contextual fragmentation of iterative sliding-window decoding
used by RankGPT (window size 20, stride 10).
Full hyperparameters ($\beta$, $\alpha$, $\lambda$, $\tau$, learning rates, hardware) are in Appendix~\ref{sec:complexity_appendix}.

\begin{table}[t]
\centering
\small
\caption{Ablation study of the HeadRank framework (Qwen3-0.6B). \textbf{Ours} utilizes Top-8 core heads, ALPS data construction, and the full joint DPO objective. NQ and FiQA are evaluated with NDCG@10; 2Wiki and Hotpot with Recall@5.}
\label{tab:ablation_compact}
\setlength{\tabcolsep}{4pt}
\begin{tabular}{@{} l | cccc @{}}
\toprule
\textbf{Method} & \textbf{NQ} & \textbf{FiQA} & \textbf{2Wiki} & \textbf{Hotpot} \\
\midrule
\cellcolor{ourscolor}\textbf{Ours (Full Framework)} & \cellcolor{ourscolor}\textbf{43.25} & \cellcolor{ourscolor}\textbf{25.99} & \cellcolor{ourscolor}\textbf{72.12} & \cellcolor{ourscolor}\textbf{85.25} \\
\midrule
\quad \textit{w/ SFT (RankNet)} & 37.39 & 22.92 & 69.65 & 79.60 \\
\quad \textit{w/ all heads} & 20.52 & 8.97 & 67.05 & 54.80 \\
\quad \textit{w/o Entropy Gating} $G_{\text{ent}}$ & 39.50 & 24.12 & 69.80 & 82.15 \\
\quad \textit{w/o ALPS strategy} & 40.85 & 23.50 & 70.45 & 83.20 \\
\quad \textit{w/o regularizer} $\Omega(\bm{s}_\theta)$ & 38.60 & 23.95 & 68.20 & 80.50 \\
\quad \textit{w/o proximal penalty} $\mathcal{L}_{\text{prox}}$ & 36.45 & 21.30 & 68.85 & 80.10 \\
\bottomrule
\end{tabular}
\end{table}

\subsection{In-Domain and Out-of-Domain Retrieval.}
Table \ref{tab:ndcg10_top40} presents NDCG@10 performance across two in-domain TREC-DL benchmarks and nine out-of-domain datasets (including BEIR subsets, NQ, and FEVER). HeadRank achieves the highest average NDCG@10 across all Qwen3 model scales (0.6B, 1.7B, and 4B), outperforming generative and decoding-free baselines as well as BM25.

For in-domain alignment on TREC-DL19 and TREC-DL20, HeadRank at 4B matches or exceeds every baseline, including the generative upper bound RankGPT. This demonstrates that projecting preference constraints into continuous attention space sharpens discriminability in long contexts.

HeadRank generalizes robustly out-of-domain despite training on only 211 MS MARCO queries. Combining ALPS and $\Omega(\bm{s}_\theta)$ yields consistent gains across medicine (TREC-COVID), encyclopedias (DBPedia), and fact-checking (FEVER). Under Qwen3-1.7B, HeadRank reaches 43.32 on NQ, substantially outperforming zero-shot attention baselines like ICR (37.09) and NIAH (35.98) that lack explicit preference alignment. This suggests the model learns generalizable semantic relevance rather than overfitting.

On a few benchmarks, generative methods retain an edge: RankGPT (4B) outperforms HeadRank on News (+3.1) and DBPedia (+4.1), likely because these knowledge-grounded datasets benefit from full autoregressive reasoning. Similarly, CoRe marginally leads on FiQA at 1.7B (+0.38). However, these per-dataset advantages do not transfer to the aggregate: HeadRank leads in overall average NDCG@10 at every scale (44.56 vs.\ 41.14 at 0.6B; 46.05 vs.\ 42.58 at 1.7B; 46.73 vs.\ 44.89 at 4B), confirming that attention-space alignment provides more \emph{consistent} cross-domain gains.

Moreover, all attention-based methods, including HeadRank, achieve a 100\% formatting success rate by construction, whereas RankGPT drops to 76.6\% SR on Qwen3-0.6B due to malformed autoregressive outputs.

\begin{figure*}[t]
  \centering
  \begin{subfigure}[t]{0.333\textwidth}
    \phantomsection
    \includegraphics[width=\linewidth]{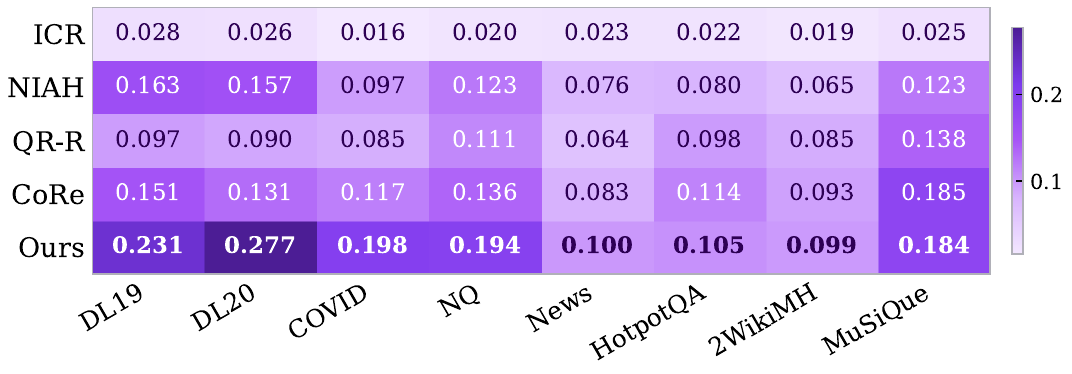}
    \caption{Qwen3-0.6B}
    \label{fig:heatmap_06b}
  \end{subfigure}%
  \hfill
  \begin{subfigure}[t]{0.333\textwidth}
    \phantomsection
    \includegraphics[width=\linewidth]{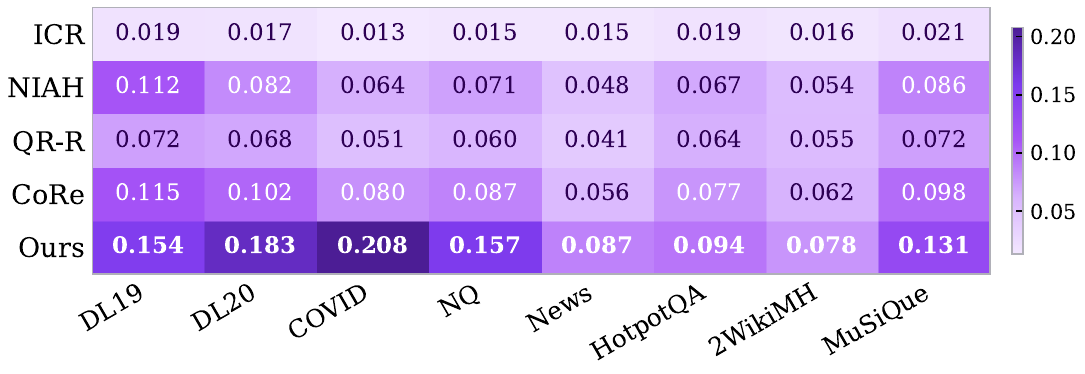}
    \caption{Qwen3-1.7B}
    \label{fig:heatmap_17b}
  \end{subfigure}%
  \hfill
  \begin{subfigure}[t]{0.333\textwidth}
    \phantomsection
    \includegraphics[width=\linewidth]{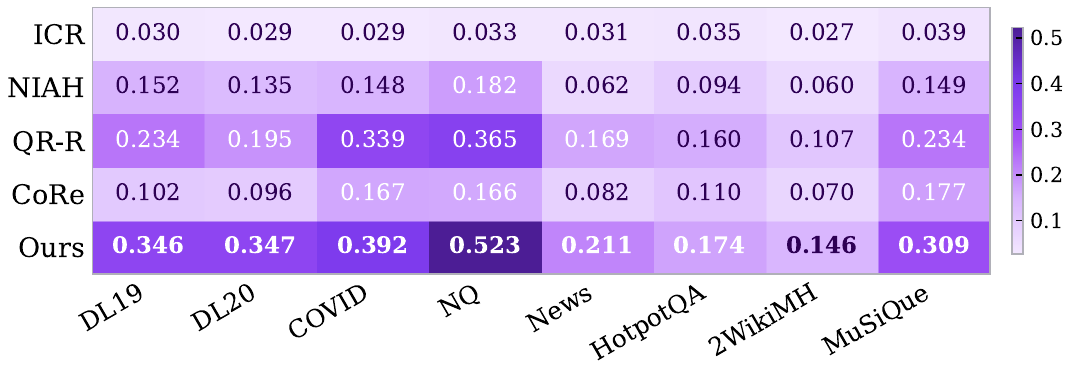}
    \caption{Qwen3-4B}
    \label{fig:heatmap_4b}
  \end{subfigure}
  \caption{%
    \textbf{Middle-zone normalized attention-score standard deviation} ($\uparrow$ better)
    across five methods, eight datasets, and three model scales.
    Lighter cells indicate more severe attention homogenization.%
  }
  \label{fig:heatmap_triad}
  \vspace{-1em}
\end{figure*}
\subsection{Multihop Reasoning Capabilities.}
Table \ref{tab:multihop} presents the results for knowledge-intensive multihop reasoning tasks (HotpotQA, 2Wiki, and MuSiQue). Unlike single-hop retrieval, these tasks require the reranker to identify multiple complementary evidence documents whose relevance emerges only through cross-document inference chains---a setting where attention-based scoring can capture inter-document bridging entities that generative methods must reason over token-by-token.
HeadRank consistently improves retrieval across all Qwen3 model scales, achieving the highest average recall. Specifically, utilizing the Qwen3-4B model, our method obtains an average Recall@2 of 59.64 and Recall@5 of 71.77. Notably, this 4B model outperforms larger generative baselines---Llama-3.1 8B (56.60 R@2) and GPT-3.5 Turbo (57.40 R@2) running RankGPT---despite being roughly half their size. Furthermore, complex multihop contexts severely disrupt the formatting stability of generative methods, as evidenced by Mistral 7B dropping to a 20.5\% Success Rate (SR). In contrast, HeadRank guarantees a 100\% SR, confirming the absolute robustness of the decoding-free alignment paradigm.

\subsection{Ablation Studies}
\label{subsec:ablation}

To dissect HeadRank's components, we conduct ablations on the Qwen3-0.6B architecture (Table \ref{tab:ablation_compact}). 

\textbf{Core Heads \& Entropy Gating.} Removing core head isolation (\textit{w/ all heads}) causes catastrophic degradation (e.g., FiQA drops from 25.99 to 8.97), proving global attention is heavily contaminated by noise. Omitting entropy gating (\textit{w/o} $G_{\text{ent}}$) consistently decreases performance, confirming that purely contrastive selection isolates unstable, high-entropy heads.

\textbf{ALPS \& Distribution Regularizer.} Replacing ALPS with random pairs (\textit{w/o ALPS}) degrades accuracy, validating that adjacent-level contrasting is essential for fine-grained discrimination. Removing the distribution regularizer (\textit{w/o} $\Omega(\bm{s}_\theta)$) causes a clear drop on long-context multihop tasks, confirming its role in preventing score homogenization among middle-ranked candidates.

\textbf{Proximal Penalty.} Finally, discarding the proximal constraint (\textit{w/o} $\mathcal{L}_{\text{prox}}$) causes representation collapse and catastrophic forgetting, severely impairing zero-shot out-of-domain generalization on NQ and FiQA.

\textbf{Attention-based DPO vs.\ Standard SFT.} Replacing DPO with pairwise SFT (\textit{w/ SFT}, RankNet-style $\sigma(s_c - s_r)$) consistently degrades all four evaluation sets. Without the frozen reference penalty $\mathcal{L}_{\text{prox}}$, unconstrained SFT aggressively fits hard targets, distorting the pre-trained attention topology and triggering representation collapse; the proximal anchor preserves linguistic priors while steering toward relevance preferences.
\begin{figure}[t]
\centering
\includegraphics[width=1.0\columnwidth]{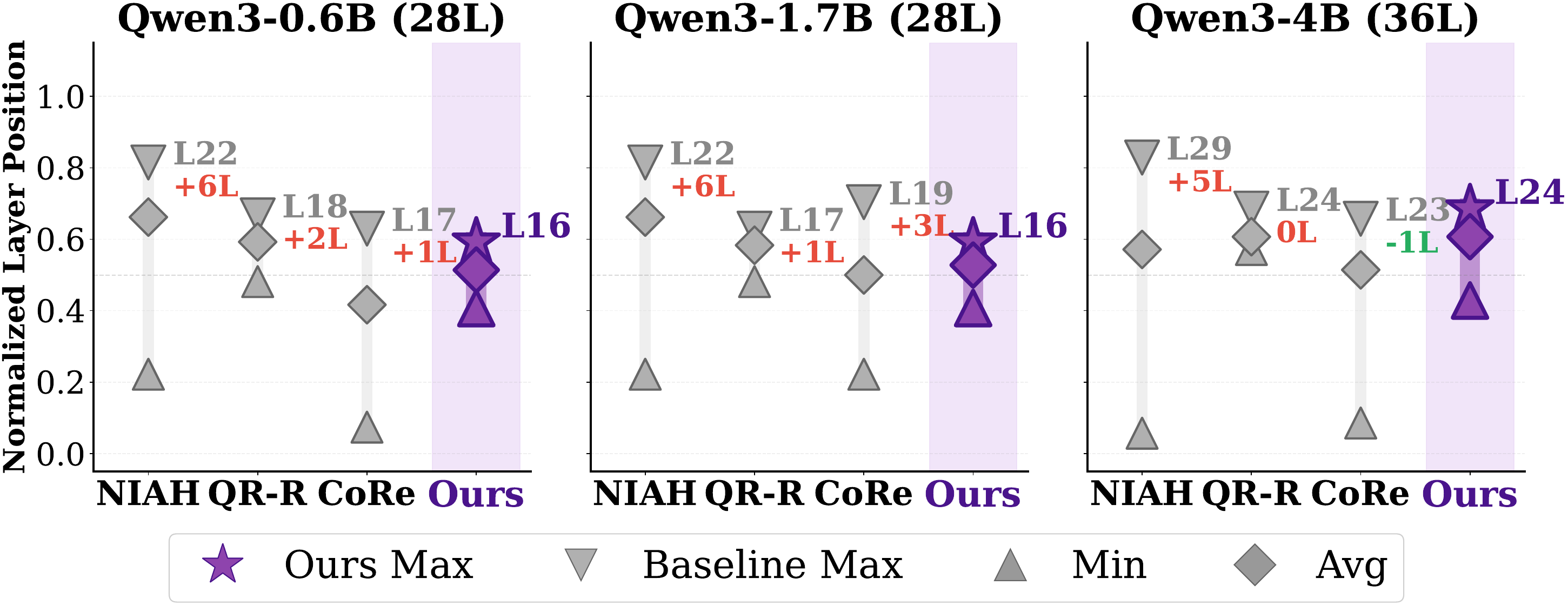}
\caption{%
    \textbf{Depth distribution of selected core heads} (Qwen3-0.6B).
    Heads above the dashed line at $l_{\max}$ are pruned for early-exit inference.%
  }
\label{fig:efficient}
\vspace{-1em}
\end{figure}

\subsection{Deep Analysis}
\label{subsec:position_variance}

\begin{figure*}[t]
  \centering
  \begin{subfigure}[t]{0.333\textwidth}
    \includegraphics[width=\linewidth]{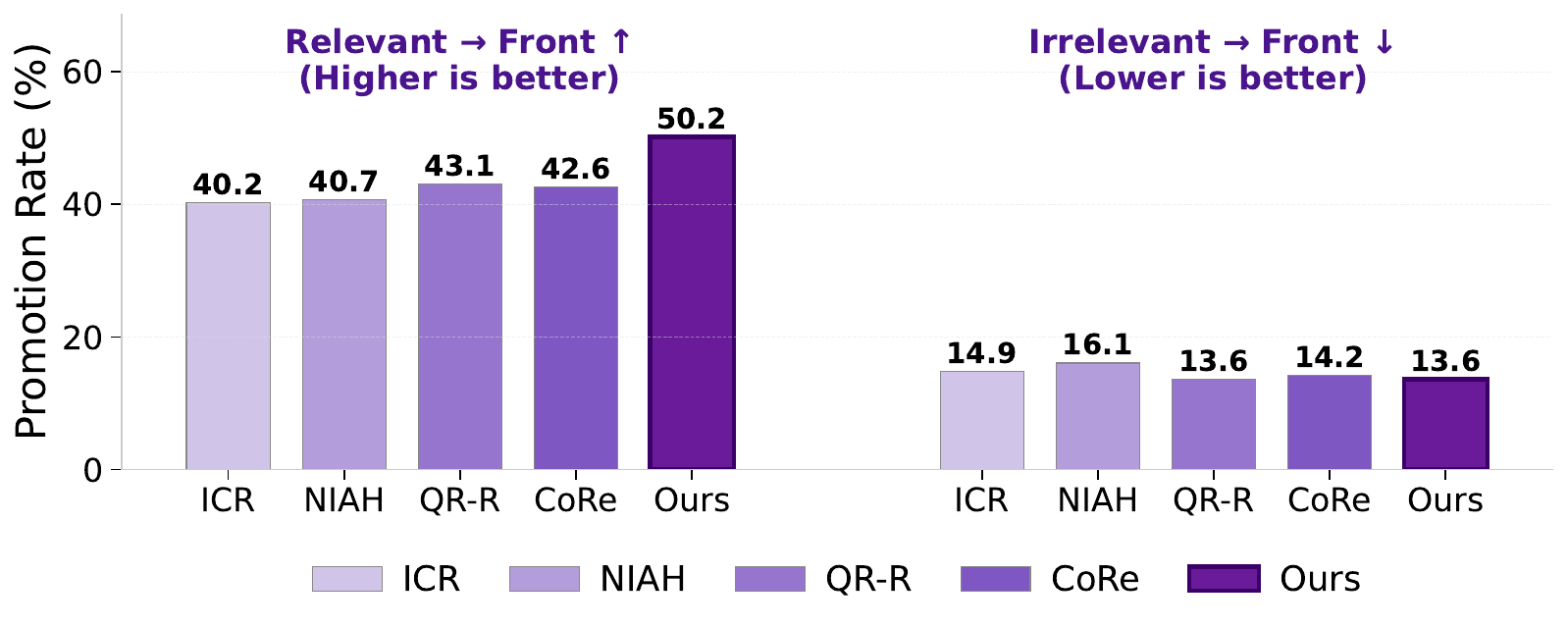}
    \caption{Qwen3-0.6B}
    \label{fig:movement_06b}
  \end{subfigure}%
  \hfill
  \begin{subfigure}[t]{0.333\textwidth}
    \includegraphics[width=\linewidth]{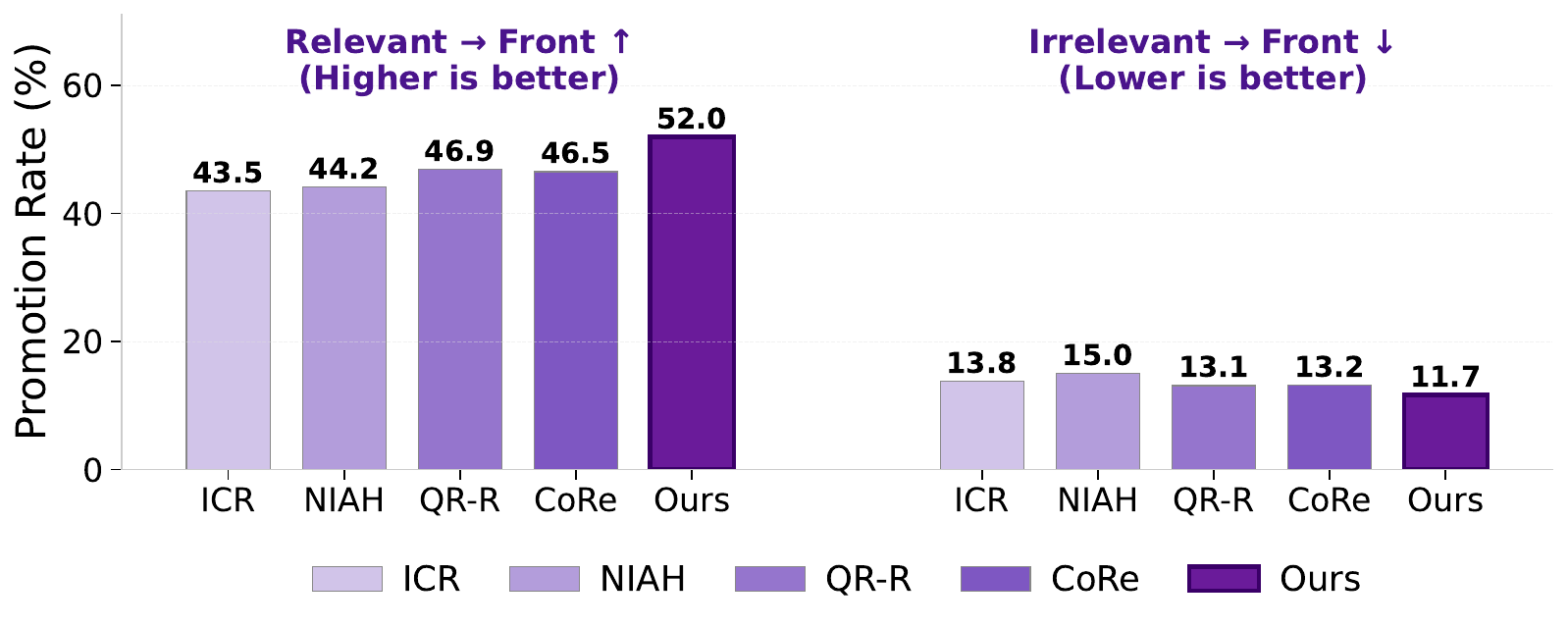}
    \caption{Qwen3-1.7B}
    \label{fig:movement_17b}
  \end{subfigure}%
  \hfill
  \begin{subfigure}[t]{0.333\textwidth}
    \includegraphics[width=\linewidth]{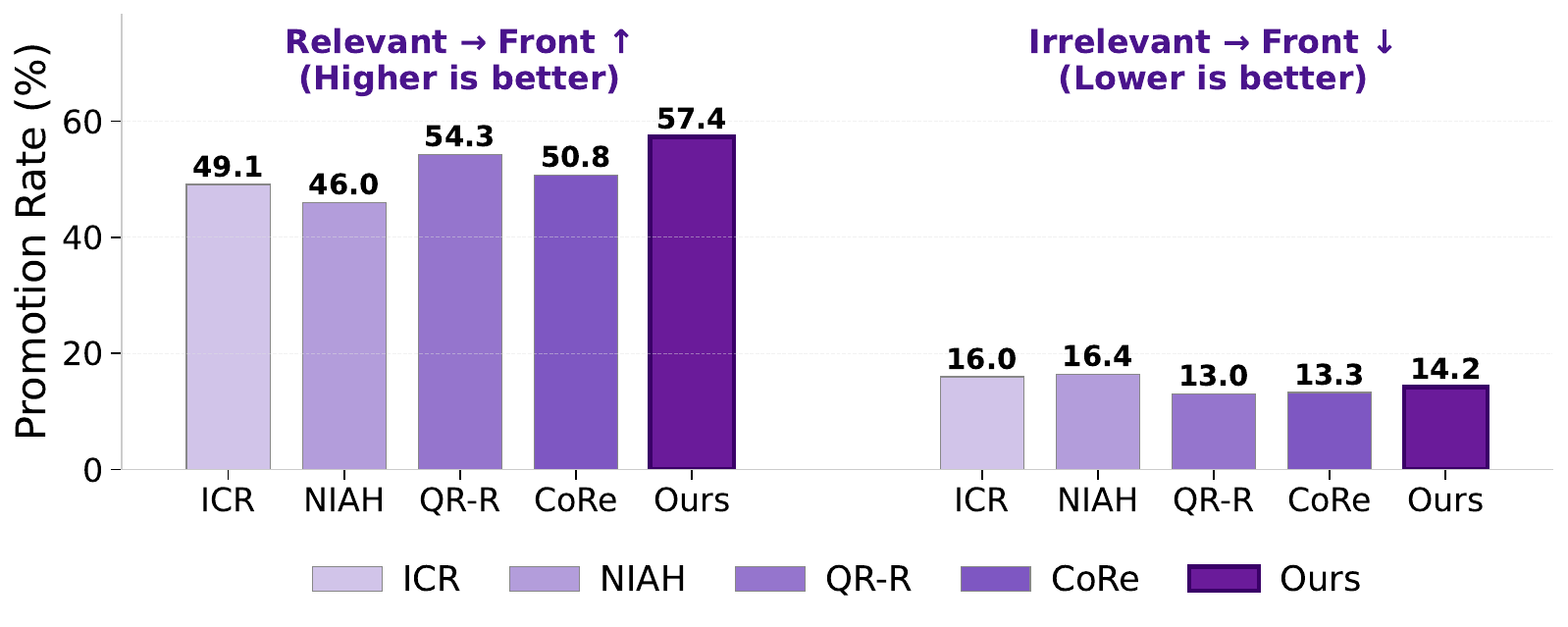}
    \caption{Qwen3-4B}
    \label{fig:movement_4b}
  \end{subfigure}
  \caption{%
    \textbf{Middle-to-front promotion rates} averaged across eight datasets.
    Documents in the middle zone (25\textsuperscript{th}--75\textsuperscript{th} percentile of BM25 ranks)
    are checked for promotion to the top quartile after reranking.
    Left bars: relevant documents promoted ($\uparrow$ better);
    right bars: irrelevant documents promoted ($\downarrow$ better).%
  }
  \label{fig:movement}
\end{figure*}

\begin{table*}[t]
\centering
\scriptsize
\caption{Token-level attention visualization for FEVER. Claim: ``Jeb Bush's mother is Laura Bush.'' Deeper \hlc[violet!80!white]{purple} indicates higher attention weight. BM25/ICR/RankGPT columns show the rank assigned by each method.}
\label{tab:attn_fever_194182}
\setlength{\tabcolsep}{2pt}
\begin{tabular}{c p{0.68\textwidth} ccc c}
\toprule
\textbf{Rank} & \textbf{Passage (token-level attention coloring)} & \textbf{BM25} & \textbf{ICR} & \textbf{RankGPT} & \textbf{Rel.} \\
\midrule
\#1 & \textbf{Jeb Bush} \hlc[violet!6!white]{Jeb} \hlc[violet!9!white]{Bush} John Ellis `` '' \hlc[violet!8!white]{Bush} \hlc[violet!11!white]{Sr} \hlc[violet!5!white]{.} ( born February , \hlc[violet!11!white]{)} is an American businessman and politician who served as the \hlc[violet!6!white]{43rd} \hlc[violet!5!white]{Governor} of \hlc[violet!5!white]{Florida} \hlc[violet!6!white]{from} \hlc[violet!10!white]{to} \hlc[violet!37!white]{.} \hlc[violet!8!white]{Bush} , who grew \hlc[violet!8!white]{up} \hlc[violet!7!white]{in} \hlc[violet!9!white]{Houston} \hlc[violet!30!white]{,} \hlc[violet!7!white]{is} the \hlc[violet!74!white]{second} \hlc[violet!82!white]{son} \hlc[violet!46!white]{of} \hlc[violet!14!white]{former} President \hlc[violet!6!white]{George} H . W \ldots & 13 & 11 & 22 & \cmark \\
\addlinespace[3pt]
\#2 & \textbf{Barbara Bush} Barbara Bush nee Pierce \hlc[violet!7!white]{;} born June , ) is the wife \hlc[violet!13!white]{of} George H . W . \hlc[violet!26!white]{Bush} \hlc[violet!6!white]{,} \hlc[violet!5!white]{the} 41st President of the United States , \hlc[violet!6!white]{and} served as First Lady \hlc[violet!7!white]{of} the United States from to \hlc[violet!27!white]{.} She is the \hlc[violet!82!white]{mother} \hlc[violet!54!white]{of} George W . \hlc[violet!8!white]{Bush} \hlc[violet!17!white]{,} \ldots & 1 & 16 & 3 & -- \\
\addlinespace[3pt]
\#3 & \textbf{Political positions of Jeb Bush} Political \hlc[violet!7!white]{positions} \hlc[violet!6!white]{of} \hlc[violet!37!white]{Jeb} \hlc[violet!68!white]{Bush} \hlc[violet!100!white]{is} \hlc[violet!11!white]{a} \hlc[violet!7!white]{Republican} politician in the United States \hlc[violet!8!white]{.} \hlc[violet!7!white]{Bush} was \hlc[violet!20!white]{governor} \hlc[violet!14!white]{of} \hlc[violet!6!white]{Florida} \hlc[violet!20!white]{from} \hlc[violet!8!white]{to} \hlc[violet!39!white]{.} \hlc[violet!18!white]{He} \hlc[violet!8!white]{was} \hlc[violet!10!white]{a} \hlc[violet!6!white]{candidate} for the Republican nomination for president of the United States in the election \hlc[violet!6!white]{.} & 2 & 4 & 12 & -- \\
\addlinespace[3pt]
\#4 & \textbf{Miss Beazley (dog)} \hlc[violet!7!white]{Miss} Beazley (dog ) October -- May , ) was a Scottish Terrier which belonged to former U.S . President George W . Bush and former U.S . First Lady \hlc[violet!76!white]{Laura} \hlc[violet!100!white]{Bush} \hlc[violet!8!white]{.} Miss Beazley 's father , a Scottish terrier named Clinton , was born on November , . \ldots & 9 & 1 & 18 & -- \\
\addlinespace[3pt]
\#5 & \textbf{Governor Bush} Governor Bush may refer to \hlc[violet!6!white]{:} George W . Bush ( born ) , 46th Governor of Texas ( -- ) and 43rd President of the United States . \hlc[violet!17!white]{Jeb} \hlc[violet!100!white]{Bush} \hlc[violet!51!white]{(} born \hlc[violet!5!white]{)} \hlc[violet!8!white]{,} 43rd \hlc[violet!5!white]{Governor} of \hlc[violet!9!white]{Florida} \hlc[violet!12!white]{(} \hlc[violet!12!white]{--} \hlc[violet!6!white]{)} Category \hlc[violet!7!white]{:} and name disambiguation pages & 12 & 37 & 21 & -- \\
\bottomrule
\end{tabular}
\end{table*}

\paragraph{Attention Homogenization Across Methods and Scales.}
To determine the pervasiveness of score flatlining in the middle context zone and assess whether any existing method effectively mitigates it, Figure~\ref{fig:heatmap_triad} diagnoses this phenomenon across five methods, eight datasets, and three model scales using middle-zone normalized standard deviation. The results reveal a systemic vulnerability in baseline approaches: ICR rows remain near-white throughout, indicating a complete flatlining of its scoring function. Similarly, NIAH and CoRe recover only marginal variance, while QR-R exhibits moderate but inconsistent improvements. In stark contrast, \textbf{HeadRank} consistently occupies the darkest row across all 24 experimental conditions, with its advantage widening substantially at the 4B scale. This robust discriminability is directly attributable to the $\operatorname{Var}(\bm{s}_{\text{mid}})$ regularization term within $\Omega(\bm{s}_\theta)$, which serves as a structural wedge to forcibly break attention homogenization. Furthermore, our entropy gating mechanism, $G_{\text{ent}}$, ensures that this discriminative pressure is concentrated exclusively on stable, low-entropy heads, preventing noisy heads from dominating the variance and stabilizing the optimization process. Cross-dataset universality and scaling behavior are further validated in Appendix~\ref{sec:scaling_appendix}, with per-dataset radar profiles in Appendix~\ref{sec:radar_appendix}.

\paragraph{From Variance to Ranking Precision.}
Does this structural variance actually translate into correct positional assignments? We quantify \emph{middle-to-front promotion rates}---the fraction of relevant vs.\ irrelevant documents originally in the middle zone (BM25 ranks in the 25\textsuperscript{th}--75\textsuperscript{th} percentile) that are reranked into the top quartile, averaged across eight datasets (Figure~\ref{fig:movement}). HeadRank's selectivity gap (relevant $-$ irrelevant promotion) grows monotonically with scale: $36.6$\,pp at 0.6B ($+7.1$ over QR-R), $40.4$\,pp at 1.7B (achieving the lowest irrelevant promotion rate among all methods), and $43.1$\,pp at 4B. This monotonic scaling mirrors the variance trends observed earlier, confirming that the score headroom generated by $\Omega(\bm{s}_\theta)$ translates directly into ranking precision. ALPS hard-sample pairs reinforce these dynamics by supplying contrastive signals between adjacent-grade documents within the middle zone, effectively transforming structural score spread into downstream ranking gains.

\paragraph{Dynamic Layer Pruning and Early-Exit Efficiency.}
HeadRank achieves these discriminability gains while reducing computational overhead. Figure~\ref{fig:efficient} shows that preference alignment compresses core heads into shallower layers compared to baselines (e.g., NIAH and QR-R reach $L_{22}$ in Qwen3-0.6B). By optimizing designated heads to capture relevance signals directly, HeadRank localizes discriminability to earlier stages, enabling safe truncation at $l_{\max} = 16$ (28L models) or $l_{\max} = 24$ (36L model). This early-exit inference translates directly into wall-clock speedups, placing HeadRank on the Pareto frontier of latency versus NDCG@10 (Figure~\ref{fig:speed_perf}, Appendix~\ref{sec:efficiency_appendix}).

\paragraph{Case study.}
Table~\ref{tab:attn_fever_194182} visualizes token-level attention for the FEVER claim \textit{``Jeb Bush's mother is Laura Bush''} (false---his mother is Barbara Bush).
BM25 ranks \emph{Barbara Bush} at \#1 via a term-frequency trap (``Bush'' + ``mother''), burying the gold \emph{Jeb Bush} article at \#13.
ICR ranks \emph{Miss Beazley} (a presidential pet dog) at \#1 by surface-matching ``Laura Bush,'' compounded by lost-in-the-middle positional bias.
\textbf{HeadRank} instead concentrates attention on the relational chain ``\texttt{second}\,(74)\,$\to$\,\texttt{son}\,(82)\,$\to$\,\texttt{of}\,(46),'' promoting the gold document to \#1 (NDCG@10: 1.0 vs.\ BM25\,0.08, ICR\,0.63).
Crucially, the selected heads track \emph{dependency structure} rather than surface tokens: ``second son of'' encodes a familial relation that unambiguously identifies parentage---something neither keyword matching nor unaligned attention can resolve.
Additional case studies covering seven benchmarks are in Appendix~\ref{sec:case_study_appendix}, with multi-hop examples in Appendix~\ref{sec:appendix_multihop_attn}.

\section{Conclusion}

We presented \textbf{HeadRank}, which lifts preference optimization into continuous attention space to overcome score homogenization in decoding-free rerankers.
Trained on only 211 queries, HeadRank achieves the best average ranking quality across 14 benchmarks at three Qwen3 scales (0.6B--4B) with $\mathcal{O}(1)$-pass inference, outperforming both generative and attention-based baselines---including models twice its size on multi-hop reasoning tasks.

\section*{Limitations}

\paragraph{Training-Time Efficiency.}
Although HeadRank achieves significant \emph{inference} speedup through depth truncation, its \emph{training} cost remains comparable to full-model DPO because gradients still propagate through all layers up to $l_{\max}$.
Since all selected core heads reside within layers 11--16 (out of 28 for Qwen3-0.6B), a natural extension is to freeze layers beyond $l_{\max}$ during training, potentially reducing optimizer states and backward-pass FLOPs by over 40\%.
We leave this head-guided layer freezing strategy to future work.

\paragraph{Static Head Assignment.}
The core head set is fixed after iterative selection and shared across all queries at inference time. A query-adaptive head selection mechanism could yield further gains for heterogeneous query types but would introduce additional overhead.

\bibliography{custom}

\clearpage
\appendix

\begin{center}
  \Large\textbf{Appendix}
\end{center}
\vspace{0.5em}

\noindent\textbf{Table of Contents}
\vspace{0.3em}

\startcontents[appendix]
\printcontents[appendix]{}{1}{\setcounter{tocdepth}{2}}

\vspace{1.5em}

\section{Inference Efficiency}
\label{sec:efficiency_appendix}

\begin{figure}[h]
  \centering
  \includegraphics[width=\columnwidth]{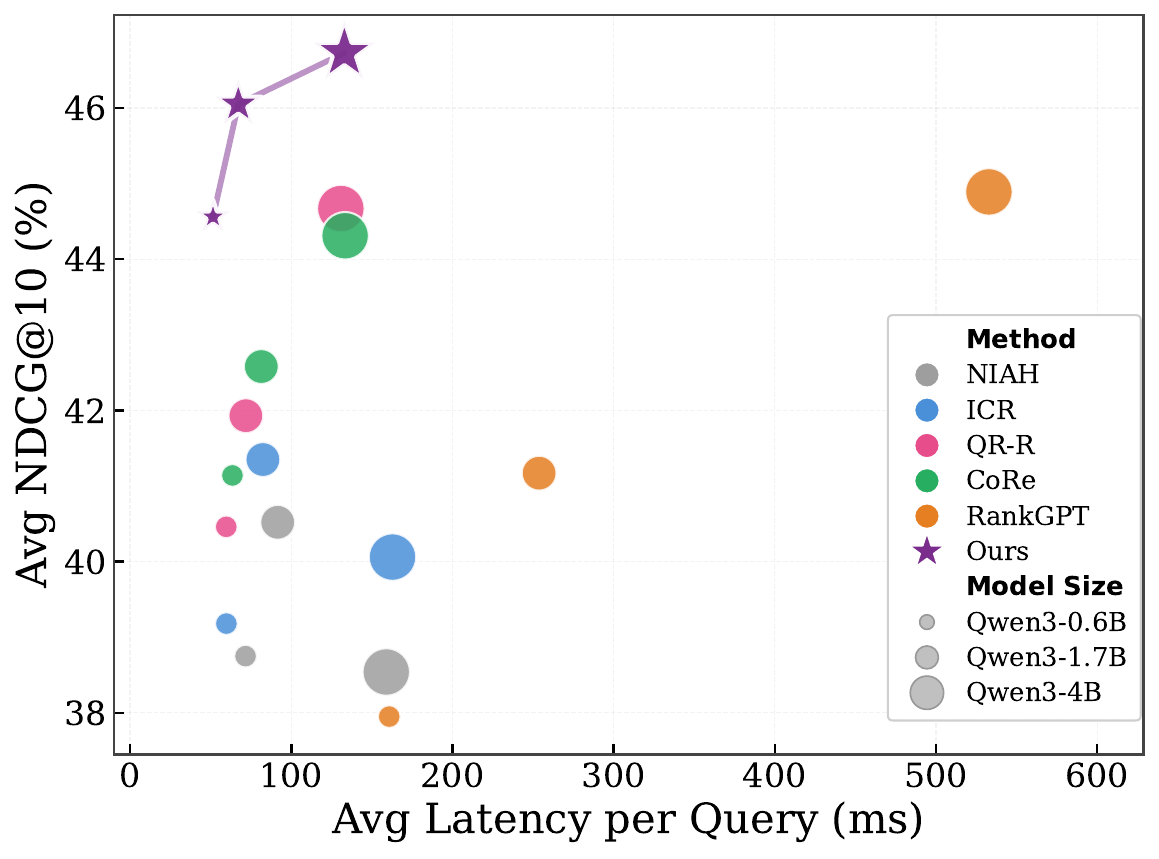}
  \caption{%
    \textbf{Inference latency vs.\ NDCG@10} across all methods and model scales.
    \textbf{HeadRank} occupies the Pareto frontier, delivering the highest
    NDCG@10 at each latency tier without generative decoding overhead.%
  }
  \label{fig:speed_perf}
\end{figure}

Figure~\ref{fig:speed_perf} plots inference latency against NDCG@10 across all
methods and model scales.
\textbf{HeadRank} occupies the Pareto frontier at every latency tier: it delivers
the highest NDCG@10 among all compared methods while avoiding the autoregressive
decoding overhead of RankGPT.
Depth truncation at layer $l_{\max}$ (Figure~\ref{fig:efficient}) enables early
exit after processing only $l_{\max}/L$ of the model's layers, yielding substantial
wall-clock inference speedups that scale with model size.

\section{Per-Dataset Radar Profiles}
\label{sec:radar_appendix}

\begin{figure*}[t]
  \centering
  \begin{subfigure}[t]{0.32\textwidth}
    \centering
    \includegraphics[width=\linewidth]{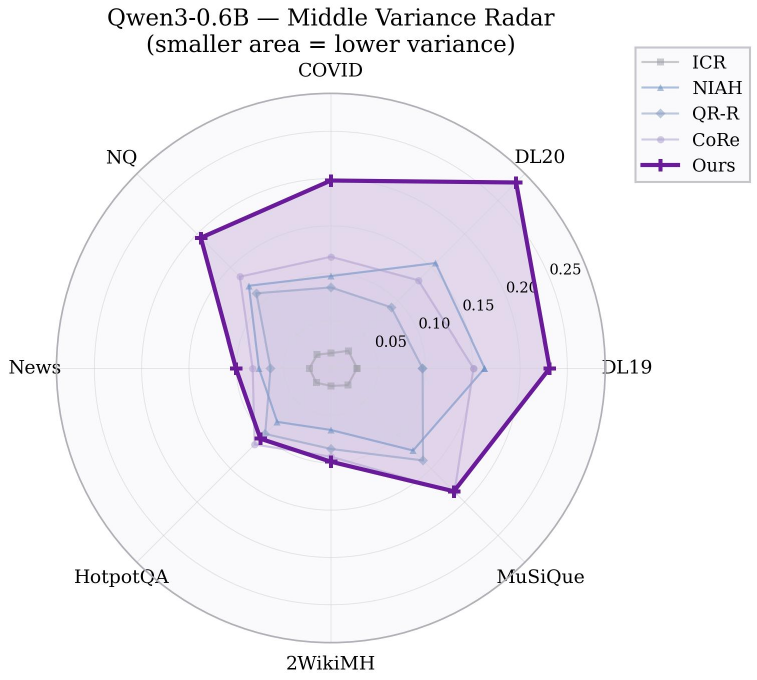}
    \caption{Qwen3-0.6B}
    \label{fig:radar_06b}
  \end{subfigure}
  \hfill
  \begin{subfigure}[t]{0.32\textwidth}
    \centering
    \includegraphics[width=\linewidth]{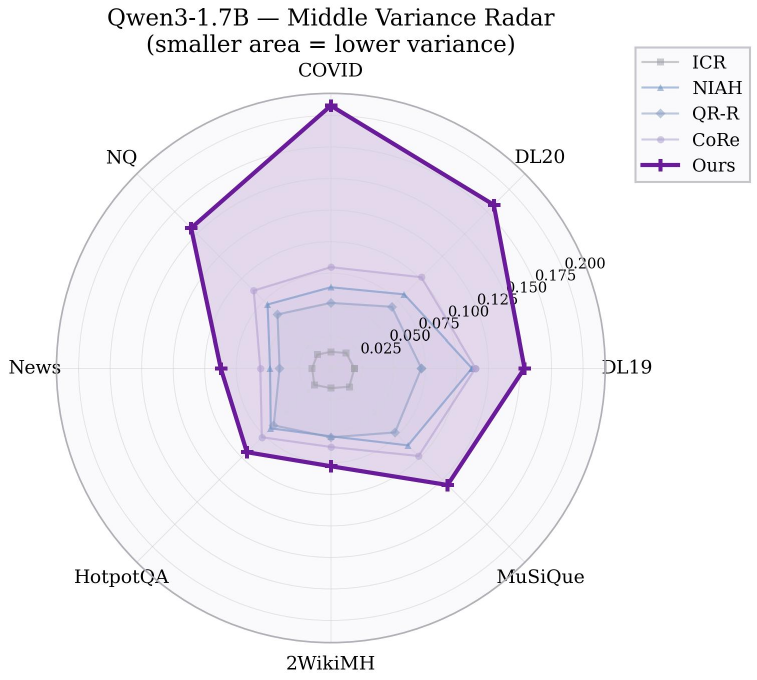}
    \caption{Qwen3-1.7B}
    \label{fig:radar_17b}
  \end{subfigure}
  \hfill
  \begin{subfigure}[t]{0.32\textwidth}
    \centering
    \includegraphics[width=\linewidth]{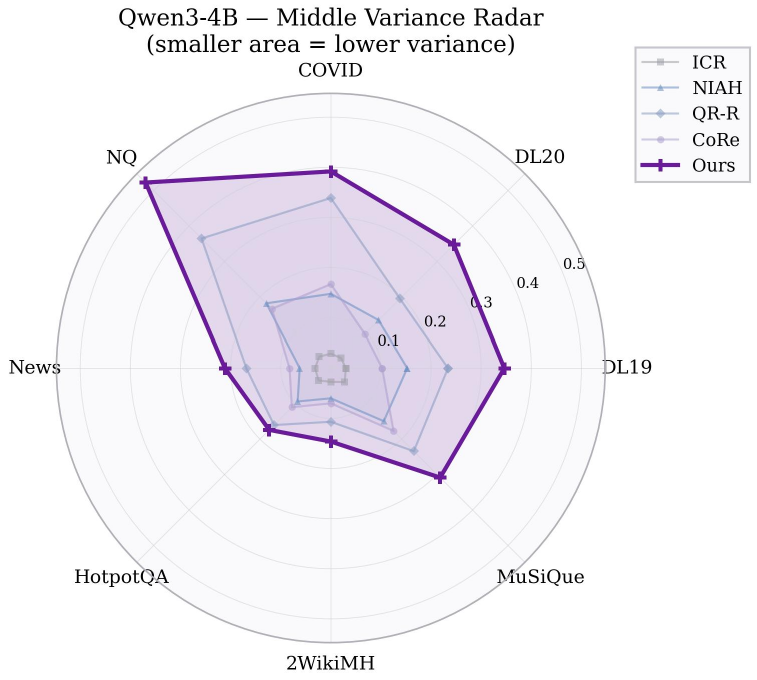}
    \caption{Qwen3-4B}
    \label{fig:radar_4b}
  \end{subfigure}
  \caption{Per-dataset radar profiles of normalised middle-zone
    standard deviation. HeadRank consistently occupies the outermost
    polygon across all three model scales.}
  \label{fig:radar}
\end{figure*}

Figure~\ref{fig:radar} presents per-dataset radar charts of normalised
middle-zone standard deviation at 0.6B, 1.7B, and 4B scales.
HeadRank traces the outermost polygon across all eight axes at both
scales, confirming that the variance gains observed in the heatmaps
(Section~\ref{subsec:position_variance}) are consistent across ad-hoc retrieval
(DL19, DL20, COVID, NQ, News) and multi-hop reasoning benchmarks
(HotpotQA, 2WikiMultiHop, MuSiQue).

\section{Cross-Dataset Universality and Scaling Behavior}
\label{sec:scaling_appendix}

\begin{figure}[h]
  \centering
  \includegraphics[width=\columnwidth]{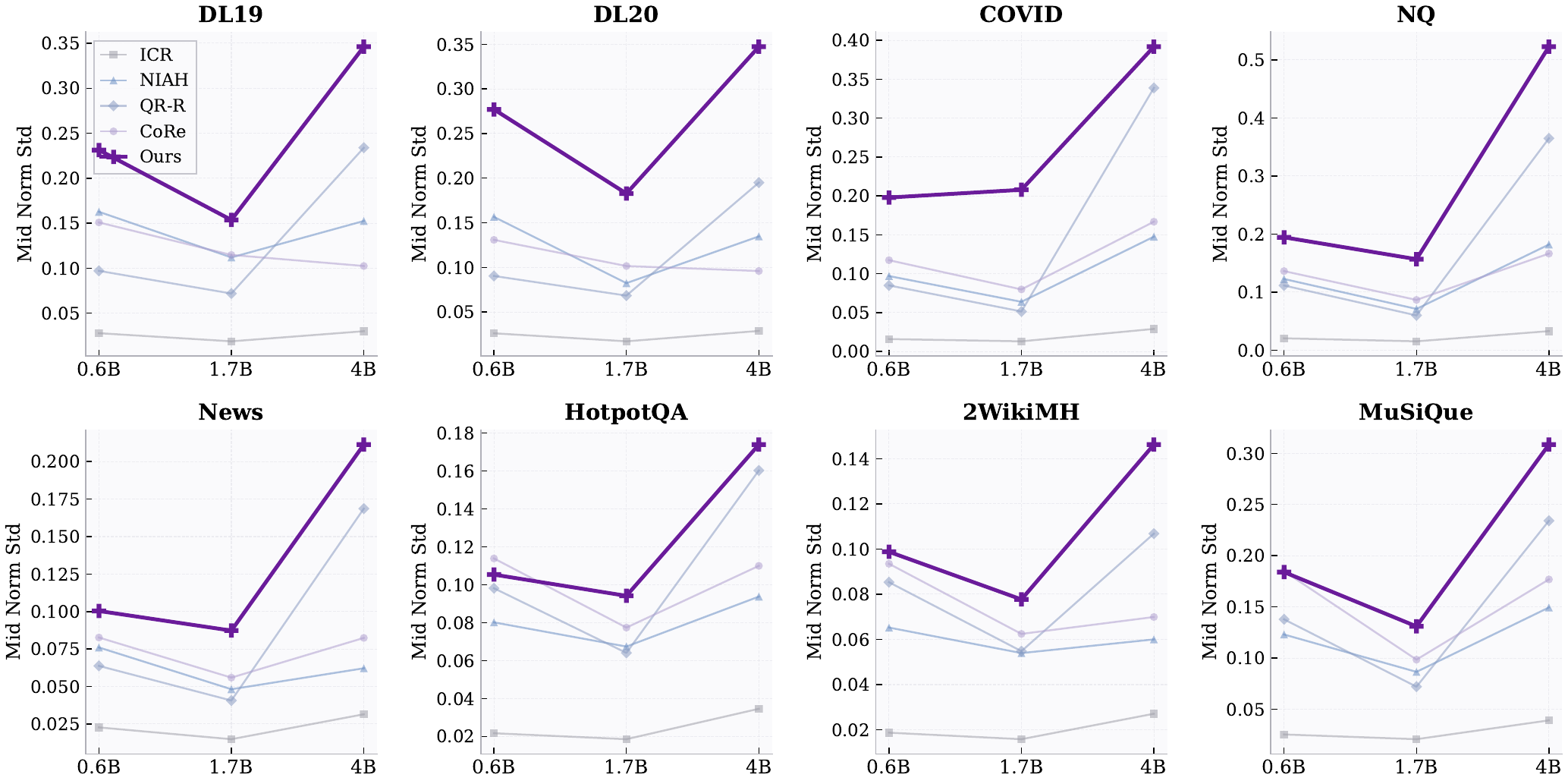}
  \caption{%
    \textbf{Scaling behavior of middle-zone normalized std} ($\uparrow$ better)
    as model capacity increases from 0.6B $\to$ 1.7B $\to$ 4B, reported
    per dataset ($2{\times}4$ grid).
    \textbf{HeadRank} exhibits monotonic improvement on six of eight
    datasets and dominates all baselines at every scale.%
  }
  \label{fig:scale_trend}
\end{figure}

Having established in Section~\ref{subsec:position_variance} that HeadRank
resolves middle-zone homogenization, a natural follow-up is whether this
advantage generalizes beyond any single benchmark and how it interacts with
model capacity.
Per-dataset radar profiles (Figure~\ref{fig:radar}) confirm that
discriminability gains are not dataset-specific artifacts: at both 0.6B and 4B,
HeadRank traces the outermost polygon across all eight radar axes---spanning
both ad-hoc retrieval and multi-hop reasoning---while ICR collapses to the
innermost ring and no baseline achieves comparable cross-dataset coverage.
Figure~\ref{fig:scale_trend} further reveals that HeadRank's middle-zone variance
increases monotonically with model size on six of eight datasets, whereas baselines
plateau or regress on multi-hop collections.
This scaling trajectory suggests that larger checkpoints surface richer
retrieval-head signal that $G_{\text{ent}}$ can isolate with increasing
precision, enabling the $\operatorname{Var}(\bm{s}_{\text{mid}})$ penalty to amplify
progressively finer score differences as model capacity grows.

\section{Hyperparameter Details}
\label{sec:complexity_appendix}

All experiments are conducted on 8\,$\times$\,NVIDIA H20 GPUs (97{,}871\,MiB VRAM each)
using the Megatron-LM distributed training framework.
Tensor parallelism is set to $\mathrm{TP}{=}4$ for the 0.6B and 1.7B scales and
$\mathrm{TP}{=}8$ for the 4B scale, with a micro-batch size of 1 per GPU and a
global batch size of 32.
Scale-specific learning rates are $1\times10^{-5}$ (0.6B), $5\times10^{-6}$ (1.7B),
and $2\times10^{-6}$ (4B).
The proximal regularisation coefficient is $\beta{=}0.05$, the contrastive loss
weight is $\alpha{=}0.05$, and attention scores are aggregated over $K{=}8$ core
heads via summation.
Gradient norms are clipped at 5.0, and all models are trained for a single epoch.
Convergence is reached at approximately 1{,}800 steps for the 0.6B model,
2{,}200 steps for the 1.7B model, and 2{,}800 steps for the 4B model.

\section{Core Head Selection Details}
\label{sec:head_appendix}

Core retrieval heads are selected by the entropy-regularized scoring function
(Eq.~\eqref{eq:entropy_selection}) with $\lambda{=}0.1$ and $\tau{=}0.001$.
We apply the procedure \emph{twice}: once on the pretrained base model (Before DPO)
and once on the final aligned checkpoint (After DPO), selecting the top-$K{=}8$
heads each time.
Table~\ref{tab:head_comparison} reports all selected heads as (Layer, Head) pairs
ranked by descending score; heads that appear in both sets are marked with
\cmark{} in the Shared column.

\begin{table*}[t]
\centering
\small
\setlength{\tabcolsep}{6pt}
\caption{Top-8 core retrieval heads selected before and after HeadRank alignment
for each Qwen3 scale, using entropy gating ($\lambda{=}0.1$, $\tau{=}0.001$).
Entries are sorted by post-DPO score.
\cmark{} marks heads appearing in both sets (highlighted in \colorbox{ourscolor!50}{\strut purple}).}
\label{tab:head_comparison}
\begin{tabular}{@{} c | ccc | ccc | ccc @{}}
\toprule
\rowcolor{ourscolor!70}
& \multicolumn{3}{c|}{\textbf{Qwen3-0.6B}}
& \multicolumn{3}{c|}{\textbf{Qwen3-1.7B}}
& \multicolumn{3}{c}{\textbf{Qwen3-4B}} \\
\rowcolor{ourscolor!40}
\textbf{Rank}
& \textbf{Before} & \textbf{After} & \textbf{Shared}
& \textbf{Before} & \textbf{After}  & \textbf{Shared}
& \textbf{Before} & \textbf{After}  & \textbf{Shared} \\
\midrule
1 & L11-H2  & L14-H14 &                              & L14-H14 & L11-H2  & \cellcolor{ourscolor!50}\cmark & L21-H8  & L24-H29 & \cellcolor{ourscolor!50}\cmark \\
2 & L11-H3  & L11-H3  & \cellcolor{ourscolor!50}\cmark & L11-H2  & L14-H6  & \cellcolor{ourscolor!50}\cmark & L21-H10 & L21-H8  & \cellcolor{ourscolor!50}\cmark \\
3 & L16-H7  & L16-H13 &                              & L11-H3  & L16-H8  &                              & L22-H4  & L21-H10 & \cellcolor{ourscolor!50}\cmark \\
4 & L21-H9  & L11-H2  & \cellcolor{ourscolor!50}\cmark & L13-H14 & L11-H3  & \cellcolor{ourscolor!50}\cmark & L23-H10 & L23-H10 & \cellcolor{ourscolor!50}\cmark \\
5 & L21-H8  & L14-H6  &                              & L14-H6  & L16-H7  &                              & L24-H29 & L22-H4  & \cellcolor{ourscolor!50}\cmark \\
6 & L6-H7   & L13-H6  &                              & L16-H14 & L16-H15 &                              & L21-H11 & L23-H30 &                              \\
7 & L17-H12 & L16-H8  &                              & L16-H13 & L14-H15 &                              & L20-H16 & L21-H18 &                              \\
8 & L13-H14 & L16-H14 &                              & L19-H5  & L16-H14 & \cellcolor{ourscolor!50}\cmark & L20-H15 & L15-H11 &                              \\
\midrule
\rowcolor{ourscolor!30}
\multicolumn{1}{c|}{\textbf{Shared}} & \multicolumn{3}{c|}{2\,/\,8}
& \multicolumn{3}{c|}{4\,/\,8}
& \multicolumn{3}{c}{5\,/\,8} \\
\bottomrule
\end{tabular}
\end{table*}

Several patterns emerge from Table~\ref{tab:head_comparison}.
\textbf{(i)~DPO consistently shifts heads toward shallower layers.}
Before alignment, selected heads span a wide range at every scale:
layers 6--21 for 0.6B, 11--19 for 1.7B, and 20--24 for 4B.
After DPO, each model \emph{compresses} its head distribution into a
tighter, shallower band---0.6B narrows from 6--21 to 11--16,
1.7B from 11--19 to 11--16 (dropping the L19 outlier and
concentrating four of eight heads in layer~16),
and 4B introduces L15-H11 while retaining five core heads in layers
21--24, effectively extending its depth floor downward by five layers.
This uniform shallowing suggests that HeadRank discovers that
lower-layer heads---closer to surface lexical features and with lower
entropy---carry cleaner ranking signals once preference-aligned.
\textbf{(ii)~Scale-dependent depth bands persist after alignment.}
Despite the consistent shallowing trend, each scale retains a
characteristic operating depth: 0.6B settles in layers 11--16,
1.7B similarly in 11--16, and 4B in 15--24.
The depth band shifts upward with model capacity,
indicating that larger models encode retrieval signal in progressively
deeper representations.
\textbf{(iii)~Stability increases with scale.}
The shared-head ratio rises from 2/8 (0.6B) to 4/8 (1.7B) and 5/8 (4B),
consistent with the hypothesis that larger models develop
well-differentiated retrieval heads during pretraining, which DPO
further sharpens without displacing.

\section{RankGPT Prompt Template}
\label{sec:prompt_appendix}

We present the zero-shot RankGPT prompt~\cite{sun2023chatgpt} used as the
generative baseline throughout our evaluation.
The prompt follows a multi-turn dialogue structure consisting of a system
prompt, iterative per-passage user turns with assistant acknowledgements,
and a closing post-prompt that requests the final ranked output.

\tcbset{
  promptstyle/.style={
    colback=violet!3!white,         
    colframe=violet!28!black,       
    colbacktitle=violet!40!black,   
    coltitle=white,
    fonttitle=\bfseries,
    boxrule=0.5pt,                  
    arc=3pt,
    left=6pt, right=6pt, top=4pt, bottom=4pt,
    toptitle=2pt, bottomtitle=2pt,
    breakable
  }
}

\begin{tcolorbox}[promptstyle, title=System Prompt]
You are RankGPT, an intelligent assistant that can rank passages based on
their relevancy to the query.
I will provide you with \texttt{[NUM]} passages, each indicated by number
identifier [].
Rank the passages based on their relevance to query: \texttt{[QUERY]}.
\end{tcolorbox}

\begin{tcolorbox}[promptstyle, title=Zero-shot User Instruction]
\textbf{Iterative User Message (per passage):}\\
\texttt{[RANK]} \texttt{[Passage Content (truncated)]}

\medskip
\textbf{Iterative Assistant Message (per passage):}
Received passage \texttt{[RANK]}.

\medskip
\textbf{Post Prompt:}
Search Query: \texttt{[QUERY]}.
Rank the \texttt{[NUM]} passages above based on their relevance to the search
query.
The passages should be listed in descending order using identifiers.
The most relevant passages should be listed first.
The output format should be \texttt{[] > []}, e.g., \texttt{[1] > [2]}.
Only provide the ranking results, do not say any word or explain.
\end{tcolorbox}

\paragraph{Sliding-window deployment.}
The prompt is deployed via a \emph{sliding window} (size 20, stride 10) over
the BM25 top-40 candidate list, sweeping from the bottom of the list upward so
that each document is ranked at least twice.
Per-window responses are parsed to extract integer identifiers; a window is
counted as a formatting success if $\geq 50\%$ of the expected identifiers are
recovered.
The same prompt is reused without modification for the multihop
(HotpotQA, 2Wiki, MuSiQue) and MLDR variants, with only the passage content
and query language changing.

\section{Token-Level Attention Visualization}
\label{sec:case_study_appendix}
\label{sec:attn_viz}

Tables~\ref{tab:attn_dbpedia_INEX_LD-2012383}--\ref{tab:attn_dl19_1106007} visualize the token-level attention distribution of \textbf{HeadRank} across seven diverse benchmarks. Each token is highlighted with a purple background whose intensity reflects its attention weight---deeper \hlc[violet!80!white]{purple} indicates higher attention. The BM25, ICR, and RankGPT columns show the rank each method assigns to the same passage for comparison.

\paragraph{Compositional entity retrieval (Table~\ref{tab:attn_dbpedia_INEX_LD-2012383}).}
The query \textit{``famous computer scientists disappeared at sea''} requires matching both a profession and a fate.
\textbf{HeadRank} places an extremely concentrated attention spike on the token \textit{``sea''} in the Jim Gray article (the sole gold document), promoting it from BM25 rank~\#4 to \#1 (NDCG@10\,=\,1.0).
In contrast, RankGPT preserves the BM25 top-3 (generic list pages) unchanged (NDCG@10\,=\,0.43), while ICR ranks an irrelevant ``Disappearance of Genette Tate'' (a missing child case) at \#1 (NDCG@10\,=\,0.63)---matching ``disappeared'' but ignoring ``computer scientists''.

\paragraph{Financial comparison reasoning (Table~\ref{tab:attn_fiqa_9565}).}
For \textit{``What are the tax benefits of dividends vs selling stock,''} attention concentrates on comparison-relevant tokens (``selling'', ``stock'', ``dividends''), enabling \textbf{HeadRank} to achieve perfect NDCG@10\,=\,1.0, versus ICR (0.31) and RankGPT (0.26).

\begin{table*}[t]
\centering
\scriptsize
\caption{Token-level attention visualization for DBPedia. Query: ``famous computer scientists disappeared at sea'' Deeper \hlc[violet!80!white]{purple} indicates higher attention weight. BM25/ICR/RankGPT columns show the rank assigned by each method.}
\label{tab:attn_dbpedia_INEX_LD-2012383}
\setlength{\tabcolsep}{2pt}
\begin{tabular}{c p{0.68\textwidth} ccc c}
\toprule
\textbf{Rank} & \textbf{Passage (token-level attention coloring)} & \textbf{BM25} & \textbf{ICR} & \textbf{RankGPT} & \textbf{Rel.} \\
\midrule
\#1 & \textbf{Jim Gray (computer scientist)} Jim Gray (computer scientist ) James Nicholas "Jim " (born January , ; lost \hlc[violet!30!white]{at} \hlc[violet!100!white]{sea} \hlc[violet!51!white]{January} , \hlc[violet!7!white]{;} declared \hlc[violet!6!white]{deceased} May , ) was an American \hlc[violet!6!white]{computer} \hlc[violet!6!white]{scientist} who received the \hlc[violet!8!white]{Turing} \hlc[violet!7!white]{Award} in "for seminal contributions to database and transaction processing res & 4 & 2 & 4 & \cmark \\
\addlinespace[3pt]
\#2 & \textbf{Computer scientist} \hlc[violet!37!white]{Computer} \hlc[violet!15!white]{scientist} A \hlc[violet!9!white]{computer} \hlc[violet!27!white]{is} \hlc[violet!8!white]{a} scientist \hlc[violet!41!white]{who} has acquired knowledge of computer \hlc[violet!48!white]{science} \hlc[violet!29!white]{,} the study of the theoretical foundations of information and \hlc[violet!7!white]{computation} \hlc[violet!8!white]{and} \hlc[violet!31!white]{their} \hlc[violet!100!white]{application.Computer} \hlc[violet!64!white]{scientists} \hlc[violet!63!white]{typically} \hlc[violet!11!white]{work} on the theoretical \hlc[violet!8!white]{side} of computer \hlc[violet!41!white]{systems} \hlc[violet!15!white]{,} as oppo & 5 & 5 & 5 & -- \\
\addlinespace[3pt]
\#3 & \textbf{Albert Lindsey Zobrist} Albert Lindsey Zobrist \hlc[violet!7!white]{(born} February , \hlc[violet!6!white]{)} \hlc[violet!10!white]{is} \hlc[violet!5!white]{an} American \hlc[violet!13!white]{computer} \hlc[violet!80!white]{scientist} \hlc[violet!100!white]{,} \hlc[violet!9!white]{games} \hlc[violet!24!white]{researcher} \hlc[violet!9!white]{,} \hlc[violet!7!white]{and} \hlc[violet!12!white]{inventor} \hlc[violet!36!white]{of} \hlc[violet!8!white]{the} \hlc[violet!52!white]{famous} \hlc[violet!68!white]{Zobrist} \hlc[violet!20!white]{Hashing} \hlc[violet!19!white]{,} \hlc[violet!22!white]{which} was \hlc[violet!15!white]{published} \hlc[violet!5!white]{in} \hlc[violet!51!white]{.} \hlc[violet!14!white]{He} is \hlc[violet!59!white]{further} \hlc[violet!7!white]{author} \hlc[violet!7!white]{of} \hlc[violet!5!white]{the} \hlc[violet!10!white]{first} Go \hlc[violet!15!white]{program} \hlc[violet!29!white]{in} as \hlc[violet!6!white]{part} of his \hlc[violet!11!white]{PhD} \hlc[violet!16!white]{Thesis} \hlc[violet!24!white]{on} patte & 10 & 19 & 10 & -- \\
\addlinespace[3pt]
\#4 & \textbf{Disappearance of Genette Tate} \hlc[violet!5!white]{Disappearance} \hlc[violet!20!white]{of} Genette \hlc[violet!16!white]{Tate} Louise (5 May -- \hlc[violet!28!white]{disappeared} August ) is an English girl whose \hlc[violet!7!white]{disappearance} became a \hlc[violet!8!white]{famous} missing person case when she went missing \hlc[violet!100!white]{at} \hlc[violet!11!white]{age} while delivering newspapers in Aylesbeare , Devon , England . & 27 & 1 & 26 & -- \\
\addlinespace[3pt]
\#5 & \textbf{Douglas T. Ross} Douglas T . Ross Taylor "Doug " (21 December -- January ) was an American computer \hlc[violet!12!white]{scientist} \hlc[violet!100!white]{pioneer} \hlc[violet!7!white]{,} and Chairman of SofTech , Inc . He is \hlc[violet!8!white]{most} famous for \hlc[violet!7!white]{originating} \hlc[violet!44!white]{the} \hlc[violet!21!white]{term} \hlc[violet!6!white]{CAD} for \hlc[violet!9!white]{computer-aided} design , and is considered to be the \hlc[violet!13!white]{father} \hlc[violet!8!white]{of} \hlc[violet!40!white]{Automatically} \hlc[violet!10!white]{Prog} & 11 & 13 & 11 & -- \\
\bottomrule
\end{tabular}
\end{table*}

\begin{table*}[t]
\centering
\scriptsize
\caption{Token-level attention visualization for FiQA. Query: ``What are the tax benefits of dividends vs selling stock'' Deeper \hlc[violet!80!white]{purple} indicates higher attention weight. BM25/ICR/RankGPT columns show the rank assigned by each method.}
\label{tab:attn_fiqa_9565}
\setlength{\tabcolsep}{2pt}
\begin{tabular}{c p{0.68\textwidth} ccc c}
\toprule
\textbf{Rank} & \textbf{Passage (token-level attention coloring)} & \textbf{BM25} & \textbf{ICR} & \textbf{RankGPT} & \textbf{Rel.} \\
\midrule
\#1 & "In the US , \hlc[violet!13!white]{dividends} are presently \hlc[violet!16!white]{taxed} at the same rates as capital \hlc[violet!8!white]{gains} \hlc[violet!8!white]{,} however \hlc[violet!50!white]{selling} \hlc[violet!100!white]{stock} \hlc[violet!74!white]{could} \hlc[violet!7!white]{lead} \hlc[violet!28!white]{to} \hlc[violet!11!white]{less} \hlc[violet!16!white]{tax} \hlc[violet!35!white]{owed} \hlc[violet!15!white]{for} \hlc[violet!9!white]{the} \hlc[violet!11!white]{same} \hlc[violet!14!white]{amount} of \hlc[violet!15!white]{cash} \hlc[violet!20!white]{raised} \hlc[violet!47!white]{,} \hlc[violet!9!white]{because} \hlc[violet!12!white]{you} \hlc[violet!6!white]{are} \hlc[violet!22!white]{getting} a \hlc[violet!16!white]{return} \hlc[violet!6!white]{of} \hlc[violet!13!white]{basis} \hlc[violet!17!white]{or} \hlc[violet!7!white]{can} \hlc[violet!20!white]{elect} \hlc[violet!8!white]{to} engage in a \hlc[violet!6!white]{""loss} \hlc[violet!9!white]{harvesting""} \ldots & 4 & 3 & 18 & \cmark \\
\addlinespace[3pt]
\#2 & The \hlc[violet!10!white]{benefit} \hlc[violet!6!white]{is} not in \hlc[violet!14!white]{taxes} . When you \hlc[violet!43!white]{sell} \hlc[violet!49!white]{a} portion of your \hlc[violet!27!white]{stock} \hlc[violet!34!white]{,} you no \hlc[violet!7!white]{longer} have a portion of your stock \hlc[violet!6!white]{.} When you get a \hlc[violet!16!white]{dividend} \hlc[violet!15!white]{,} you \hlc[violet!6!white]{still} have a portion of your stock \hlc[violet!20!white]{.} \hlc[violet!5!white]{Dividends} \hlc[violet!19!white]{are} \hlc[violet!100!white]{distributed} \hlc[violet!7!white]{from} \hlc[violet!5!white]{the} \hlc[violet!9!white]{net} \hlc[violet!25!white]{profits} \hlc[violet!5!white]{of} a \ldots & 6 & 12 & 4 & \cmark \\
\addlinespace[3pt]
\#3 & Not sure how this has got this far with no obvious discussion about the huge tax \hlc[violet!48!white]{advantages} \hlc[violet!100!white]{of} \hlc[violet!5!white]{share} buy \hlc[violet!5!white]{backs} \hlc[violet!11!white]{vs} \hlc[violet!18!white]{dividend} \hlc[violet!6!white]{paying} . \hlc[violet!10!white]{Companies} face a very simple choice with excess capital - pay to shareholders in the form of a taxable \hlc[violet!6!white]{dividend} , invest in future growth \ldots & 36 & 1 & 23 & -- \\
\addlinespace[3pt]
\#4 & Assuming a USA taxable account : \hlc[violet!6!white]{Withdrawing} funds from a brokerage account has nothing to do with taxes . Taxes are owed on the profit when you \hlc[violet!31!white]{sell} a \hlc[violet!13!white]{stock} \hlc[violet!14!white]{,} no matter what you do with the funds \hlc[violet!9!white]{.} Taxes are owed on any \hlc[violet!100!white]{dividends} \hlc[violet!66!white]{the} \hlc[violet!8!white]{stock} \hlc[violet!19!white]{produces} \hlc[violet!96!white]{,} \ldots & 18 & 36 & 14 & -- \\
\addlinespace[3pt]
\#5 & "They are \hlc[violet!22!white]{similar} in the sense that they are transferring money from the company to shareholders , but that's about it . \hlc[violet!7!white]{There} is different \hlc[violet!100!white]{tax} \hlc[violet!38!white]{treatment} , yes , but that's because they are \hlc[violet!6!white]{fundamentally} \hlc[violet!5!white]{different} . \hlc[violet!33!white]{Dividends} \hlc[violet!11!white]{transfer} \hlc[violet!13!white]{money} \hlc[violet!11!white]{equally} \hlc[violet!5!white]{to} \hlc[violet!6!white]{all} \hlc[violet!7!white]{shareholders} \hlc[violet!17!white]{,} \hlc[violet!8!white]{but} that also \hlc[violet!6!white]{reduces} \ldots & 5 & 4 & 3 & -- \\
\bottomrule
\end{tabular}
\end{table*}

\subsection{Additional Token-Level Attention Visualizations}
\label{sec:appendix_attn}

We present additional token-level attention visualizations across five diverse benchmarks, each illustrating a distinct failure mode that DPO-aligned attention overcomes.

\paragraph{Precise counting and entity disambiguation (Table~\ref{tab:attn_dl20_1131069}).}
For the query \textit{``how many sons robert kraft has,''} the gold passage states ``They had \textbf{four sons}:'' followed by their names.
Our attention peaks sharply at \texttt{four}\,(20)$\to$\texttt{sons}\,(100)$\to$\texttt{:}\,(52), precisely locating the numeric answer, then traces each son's name with moderate weights (Jonathan, Daniel, Joshua Kraft).
A confusing passage about a \emph{different} Kraft family member (``two sons and a daughter'') also activates \texttt{sons}\,(51), but the model distinguishes it via the differing count.
BM25 buries the gold passage at rank~\#24; our method promotes it to \#1 (NDCG@10: Ours\,=\,1.0 vs.\ ICR\,=\,0.41).

\paragraph{Near-synonym disambiguation: \emph{filmed} vs.\ \emph{set} (Table~\ref{tab:attn_dl20_997622}).}
The query \textit{``where is the show shameless filmed''} asks for the \emph{production location} (Los Angeles), not the fictional setting (Chicago).
In the gold passage, attention concentrates on the filming clause: \texttt{filmed}\,(67)$\to$\texttt{in}\,(100)$\to$\texttt{Los}\,(22)$\to$\texttt{Angeles}\,(33), while the setting clause ``set in Chicago'' receives minimal weight (\texttt{set}\,=\,24, \texttt{Chicago}\,=\,10).
Conversely, in the UK-version distractor (\#4), attention peaks on \texttt{is}\,(100)$\to$\texttt{set}\,(96)---the story setting rather than the filming location---correctly down-ranking it.
ICR fails catastrophically, ranking gold passages at positions 39 and 40 (NDCG@10: Ours\,=\,0.73 vs.\ ICR\,=\,0.15).

\paragraph{Medical evidence assessment (Table~\ref{tab:attn_covid_46}).}
For the biomedical query about dexamethasone as a COVID-19 treatment, attention in the top-ranked passage concentrates on the conclusion:\texttt{dexamethasone}\,(45)$\to$\allowbreak\texttt{reduce}\,(46)$\to$\allowbreak\texttt{mortality}\,(51)$\to$\allowbreak\texttt{COVID-19}\,(93)$\to$\allowbreak\texttt{patients}\,(100)$\to$\allowbreak\texttt{only}\,(51).
The notable weight on the qualifier ``\texttt{only}'' indicates the model captures the \emph{scope limitation} of the evidence (severe patients only), going beyond simple keyword matching (NDCG@10: Ours\,=\,0.89 vs.\ ICR\,=\,0.52).

\paragraph{Verb semantics in entity retrieval (Table~\ref{tab:attn_dbpedia_QALD2_tr-62}).}
For \textit{``Who created Wikipedia?,''} the \#2 passage title contains ``How A Bunch of Nobodies \textbf{Created} The World's Greatest Encyclopedia,'' where \texttt{Created} receives full attention weight (100), directly aligning with the query verb.
In the \#1 Wikipedia main page, attention shifts to the organizational creator: \texttt{Wikimedia}\,(39)$\to$\texttt{Foundation}\,(96).
BM25 ranks these gold documents at 33 and 14, overwhelmed by numerous language-edition pages that mention ``Wikipedia'' frequently (NDCG@10: Ours\,=\,1.0 vs.\ RankGPT\,=\,0.28).

\paragraph{Definition coverage (Table~\ref{tab:attn_dl19_1106007}).}
For \textit{``define visceral?,''} all top-5 passages are relevant, showcasing the model's ability to retrieve multiple valid definitions.
Attention in \#1 mirrors the query exactly: \texttt{Define}\,(100)$\to$\texttt{visceral}\,(82)$\to$\texttt{:}\,(79), while \#3 activates the medical sense \texttt{viscera)}\,(100), demonstrating polysemy-aware coverage (NDCG@10: Ours\,=\,1.0 vs.\ ICR\,=\,0.82).

\begin{table*}[t]
\centering
\scriptsize
\caption{Token-level attention visualization for DL20. Query: ``where is the show shameless filmed'' Deeper \hlc[violet!80!white]{purple} indicates higher attention weight. BM25/ICR/RankGPT columns show the rank assigned by each method.}
\label{tab:attn_dl20_997622}
\setlength{\tabcolsep}{2pt}
\begin{tabular}{c p{0.68\textwidth} ccc c}
\toprule
\textbf{Rank} & \textbf{Passage (token-level attention coloring)} & \textbf{BM25} & \textbf{ICR} & \textbf{RankGPT} & \textbf{Rel.} \\
\midrule
\#1 & Shameless (U.S . TV series ) Shameless is an American comedy-drama television series which airs on Showtime . This remake of the British series is \hlc[violet!24!white]{set} \hlc[violet!7!white]{in} \hlc[violet!10!white]{Chicago} \hlc[violet!26!white]{,} \hlc[violet!60!white]{although} \hlc[violet!67!white]{filmed} \hlc[violet!100!white]{in} \hlc[violet!22!white]{Los} \hlc[violet!33!white]{Angeles} \hlc[violet!66!white]{,} \hlc[violet!32!white]{with} \hlc[violet!17!white]{the} \hlc[violet!43!white]{exterior} \hlc[violet!40!white]{scenes} \hlc[violet!29!white]{shot} \hlc[violet!38!white]{in} \hlc[violet!31!white]{Chicago.[1} & 18 & 39 & 4 & \cmark \\
\addlinespace[3pt]
\#2 & (See \hlc[violet!7!white]{Shameless} UK TV Series). \hlc[violet!5!white]{The} \hlc[violet!13!white]{show} \hlc[violet!8!white]{first} aired on \hlc[violet!10!white]{Showtime} on January , , and is \hlc[violet!18!white]{currently} renewed for season which is set to premiere in January , \hlc[violet!10!white]{.} Although \hlc[violet!28!white]{filmed} \hlc[violet!8!white]{in} \hlc[violet!7!white]{Los} Angeles \hlc[violet!18!white]{,} \hlc[violet!34!white]{Shameless} \hlc[violet!21!white]{is} \hlc[violet!29!white]{set} \hlc[violet!15!white]{in} \hlc[violet!12!white]{ChicagoÂĢÂĻs} \hlc[violet!16!white]{Canaryville} \hlc[violet!37!white]{neighborhood} \hlc[violet!31!white]{on} the South \hlc[violet!14!white]{Side} \hlc[violet!100!white]{.} & 7 & 11 & 3 & \cmark \\
\addlinespace[3pt]
\#3 & Shameless is an American television comedy-drama which airs on Showtime . This remake of the award-winning British series is \hlc[violet!19!white]{set} \hlc[violet!15!white]{in} \hlc[violet!15!white]{Chicago} \hlc[violet!28!white]{'s} \hlc[violet!19!white]{Canaryville} \hlc[violet!100!white]{neighborhood} \hlc[violet!99!white]{on} \hlc[violet!7!white]{the} \hlc[violet!5!white]{South} \hlc[violet!35!white]{Side} \hlc[violet!69!white]{,} \hlc[violet!73!white]{although} \hlc[violet!40!white]{filmed} \hlc[violet!32!white]{in} \hlc[violet!12!white]{Los} \hlc[violet!28!white]{Angeles} \hlc[violet!13!white]{,} \hlc[violet!5!white]{with} the \hlc[violet!7!white]{exterior} \hlc[violet!13!white]{scenes} \hlc[violet!6!white]{filmed} \hlc[violet!6!white]{in} \hlc[violet!12!white]{Chicago} \hlc[violet!46!white]{.} & 40 & 40 & 31 & \cmark \\
\addlinespace[3pt]
\#4 & Shameless goes \hlc[violet!6!white]{Stateside} : The \hlc[violet!8!white]{new} U.S . \hlc[violet!6!white]{version} of the UK favourite show \hlc[violet!6!white]{premieres} tonight \hlc[violet!8!white]{on} the \hlc[violet!5!white]{Showtime} \hlc[violet!6!white]{cable} channel \hlc[violet!7!white]{.} \hlc[violet!16!white]{The} \hlc[violet!13!white]{original} : \hlc[violet!7!white]{The} \hlc[violet!6!white]{UK} version of \hlc[violet!6!white]{Shameless} \hlc[violet!100!white]{is} \hlc[violet!96!white]{set} \hlc[violet!29!white]{in} \hlc[violet!12!white]{the} \hlc[violet!48!white]{gritty} \hlc[violet!8!white]{Chatsworth} \hlc[violet!46!white]{estate} \hlc[violet!94!white]{.} \hlc[violet!40!white]{The} \hlc[violet!56!white]{role} \hlc[violet!14!white]{was} \hlc[violet!8!white]{played} \hlc[violet!5!white]{by} Anne-Marie Duff for two series in the \ldots & 31 & 12 & 6 & -- \\
\addlinespace[3pt]
\#5 & ÂĢÂĺShamelessÂĢÂĻ filming in Chicago next week \hlc[violet!5!white]{,} Extras needed \hlc[violet!46!white]{.} \hlc[violet!6!white]{According} to a new casting call , \hlc[violet!9!white]{Shameless} will return to Chicago next week \hlc[violet!7!white]{for} a five day shoot \hlc[violet!21!white]{.} The \hlc[violet!100!white]{show} \hlc[violet!48!white]{is} \hlc[violet!15!white]{primarily} \hlc[violet!7!white]{filmed} \hlc[violet!6!white]{in} Los Angeles but they spend a week or two each season \hlc[violet!10!white]{filming} \hlc[violet!5!white]{..} \hlc[violet!6!white]{According} \ldots & 4 & 4 & 14 & \cmark \\
\bottomrule
\end{tabular}
\end{table*}

\begin{table*}[t]
\centering
\scriptsize
\caption{Token-level attention visualization for COVID. Query: ``what evidence is there for dexamethasone as a treatment for COVID-19?'' Deeper \hlc[violet!80!white]{purple} indicates higher attention weight. BM25/ICR/RankGPT columns show the rank assigned by each method.}
\label{tab:attn_covid_46}
\setlength{\tabcolsep}{2pt}
\begin{tabular}{c p{0.68\textwidth} ccc c}
\toprule
\textbf{Rank} & \textbf{Passage (token-level attention coloring)} & \textbf{BM25} & \textbf{ICR} & \textbf{RankGPT} & \textbf{Rel.} \\
\midrule
\#1 & \textbf{Dexamethasone for COVID-19? Not so fast} Dexamethasone \hlc[violet!18!white]{for} \hlc[violet!43!white]{COVID-19?} \hlc[violet!9!white]{Not} so fast Recent announcements indicated without sharing any distinct published set of results \hlc[violet!6!white]{,} that the \hlc[violet!7!white]{corticosteroid} \hlc[violet!45!white]{dexamethasone} \hlc[violet!7!white]{may} \hlc[violet!46!white]{reduce} \hlc[violet!51!white]{mortality} \hlc[violet!39!white]{of} \hlc[violet!13!white]{severe} \hlc[violet!93!white]{COVID-19} \hlc[violet!100!white]{patients} \hlc[violet!51!white]{only} \hlc[violet!86!white]{.} The \hlc[violet!17!white]{recent} Coronavirus [severe acute respiratory syndrome (SARS \hlc[violet!31!white]{)} & 16 & 1 & 7 & \cmark \\
\addlinespace[3pt]
\#2 & \textbf{Effect of Dexamethasone in Hospitalized Patients with COVID-19: Preliminary Report} Effect of \hlc[violet!100!white]{Dexamethasone} \hlc[violet!21!white]{in} Hospitalized Patients \hlc[violet!6!white]{with} \hlc[violet!50!white]{COVID-19} \hlc[violet!13!white]{:} \hlc[violet!13!white]{Preliminary} \hlc[violet!5!white]{Report} \hlc[violet!13!white]{Background} \hlc[violet!8!white]{:} Coronavirus disease \hlc[violet!15!white]{(COVID-19} is \hlc[violet!35!white]{associated} diffuse \hlc[violet!10!white]{lung} \hlc[violet!23!white]{damage} \hlc[violet!50!white]{.} \hlc[violet!23!white]{Corticosteroids} \hlc[violet!15!white]{may} \hlc[violet!24!white]{modulate} \hlc[violet!6!white]{immune-mediated} \hlc[violet!31!white]{lung} \hlc[violet!42!white]{injury} \hlc[violet!55!white]{and} \hlc[violet!24!white]{reducing} \hlc[violet!30!white]{progression} to respiratory \hlc[violet!18!white]{failure} \hlc[violet!26!white]{and} deat & 5 & 5 & 3 & \cmark \\
\addlinespace[3pt]
\#3 & \textbf{Dexamethasone for COVID-19? Not so fast.} \hlc[violet!8!white]{Dexamethasone} \hlc[violet!14!white]{for} \hlc[violet!35!white]{COVID-19?} \hlc[violet!11!white]{Not} so \hlc[violet!8!white]{fast} . Recent announcements indicated without sharing \hlc[violet!6!white]{any} distinct \hlc[violet!7!white]{published} \hlc[violet!22!white]{set} of \hlc[violet!13!white]{results} \hlc[violet!6!white]{,} \hlc[violet!7!white]{that} \hlc[violet!15!white]{the} \hlc[violet!9!white]{corticosteroid} \hlc[violet!27!white]{dexamethasone} \hlc[violet!34!white]{may} \hlc[violet!100!white]{reduce} \hlc[violet!44!white]{mortality} \hlc[violet!25!white]{of} \hlc[violet!5!white]{severe} \hlc[violet!21!white]{COVID-19} \hlc[violet!61!white]{patients} \hlc[violet!65!white]{only} \hlc[violet!17!white]{.} The recent Coronavirus [severe acute respiratory syndrome (SARS & 17 & 8 & 8 & \cmark \\
\addlinespace[3pt]
\#4 & \textbf{Short-Term Corticosteroids in SARS-CoV2 Patients: Hospitalists' Perspective} Short-Term \hlc[violet!15!white]{Corticosteroids} \hlc[violet!5!white]{in} \hlc[violet!31!white]{SARS-CoV2} \hlc[violet!19!white]{Patients} \hlc[violet!9!white]{:} \hlc[violet!33!white]{Hospitalists} ' Perspective Background \hlc[violet!7!white]{:} \hlc[violet!47!white]{Dexamethasone} \hlc[violet!17!white]{a} \hlc[violet!23!white]{synthetic} glucocorticoid \hlc[violet!17!white]{has} \hlc[violet!35!white]{anti-inflammatory} \hlc[violet!34!white]{and} \hlc[violet!36!white]{immunosuppressive} \hlc[violet!100!white]{properties} \hlc[violet!37!white]{.} \hlc[violet!30!white]{There} \hlc[violet!24!white]{is} \hlc[violet!50!white]{a} \hlc[violet!66!white]{hyperinflammatory} \hlc[violet!33!white]{response} \hlc[violet!36!white]{involved} \hlc[violet!21!white]{in} \hlc[violet!19!white]{the} \hlc[violet!10!white]{clinical} \hlc[violet!18!white]{course} \hlc[violet!42!white]{of} \hlc[violet!14!white]{patients} \hlc[violet!14!white]{with} \hlc[violet!42!white]{pneumonia} & 12 & 29 & 18 & -- \\
\addlinespace[3pt]
\#5 & \textbf{Short-Term Dexamethasone in Sars-CoV-2 Patients.} \hlc[violet!6!white]{Short-Term} \hlc[violet!51!white]{Dexamethasone} \hlc[violet!5!white]{in} \hlc[violet!5!white]{Sars-CoV-2} \hlc[violet!7!white]{Patients} \hlc[violet!19!white]{.} \hlc[violet!6!white]{BACKGROUND} \hlc[violet!12!white]{a} \hlc[violet!7!white]{synthetic} glucocorticoid \hlc[violet!6!white]{has} \hlc[violet!31!white]{anti-inflammatory} \hlc[violet!15!white]{and} \hlc[violet!14!white]{immunosuppressive} \hlc[violet!28!white]{properties} \hlc[violet!15!white]{.} \hlc[violet!12!white]{There} \hlc[violet!33!white]{is} \hlc[violet!100!white]{a} \hlc[violet!46!white]{hyperinflammatory} \hlc[violet!50!white]{response} \hlc[violet!35!white]{involved} \hlc[violet!23!white]{in} \hlc[violet!9!white]{the} \hlc[violet!9!white]{clinical} \hlc[violet!26!white]{course} \hlc[violet!31!white]{of} \hlc[violet!14!white]{patients} \hlc[violet!18!white]{with} \hlc[violet!95!white]{pneumonia} \hlc[violet!35!white]{due} \hlc[violet!7!white]{to} \hlc[violet!38!white]{SARS-CoV-2} \hlc[violet!26!white]{.} \hlc[violet!5!white]{To} \hlc[violet!31!white]{date} \hlc[violet!23!white]{,} & 7 & 25 & 4 & -- \\
\bottomrule
\end{tabular}
\end{table*}

\begin{table*}[t]
\centering
\scriptsize
\caption{Token-level attention visualization for DL19. Query: ``define visceral?'' Deeper \hlc[violet!80!white]{purple} indicates higher attention weight. BM25/ICR/RankGPT columns show the rank assigned by each method.}
\label{tab:attn_dl19_1106007}
\setlength{\tabcolsep}{2pt}
\begin{tabular}{c p{0.68\textwidth} ccc c}
\toprule
\textbf{Rank} & \textbf{Passage (token-level attention coloring)} & \textbf{BM25} & \textbf{ICR} & \textbf{RankGPT} & \textbf{Rel.} \\
\midrule
\#1 & \hlc[violet!100!white]{Define} \hlc[violet!82!white]{visceral} \hlc[violet!79!white]{:} felt in or as if in the internal organs of the body \hlc[violet!5!white]{:} deep ; not intellectual : instinctive , unreasoning ÂĢÂĶ \hlc[violet!22!white]{visceral} \hlc[violet!14!white]{in} a \hlc[violet!5!white]{sentence} & 2 & 5 & 2 & \cmark \\
\addlinespace[3pt]
\#2 & Definition \hlc[violet!5!white]{of} \hlc[violet!16!white]{visceral} \hlc[violet!100!white]{.} \hlc[violet!6!white]{:} felt in \hlc[violet!6!white]{or} as if in the internal organs of the body \hlc[violet!11!white]{:} \hlc[violet!7!white]{deep} a \hlc[violet!15!white]{visceral} \hlc[violet!8!white]{conviction} \hlc[violet!19!white]{.} \hlc[violet!6!white]{:} \hlc[violet!8!white]{not} intellectual \hlc[violet!6!white]{:} instinctive , unreasoning \hlc[violet!5!white]{visceral} \hlc[violet!10!white]{drives} \hlc[violet!19!white]{.} : dealing with crude or elemental emotions : earthy a \hlc[violet!10!white]{visceral} novel \hlc[violet!8!white]{.} : of , relating \ldots & 22 & 19 & 8 & \cmark \\
\addlinespace[3pt]
\#3 & The reason \hlc[violet!20!white]{is} that there are two different kinds of fat in your belly and visceral \hlc[violet!13!white]{fat} is only one of them . \hlc[violet!6!white]{In} a health setting , the word \hlc[violet!15!white]{visceral} \hlc[violet!58!white]{means} \hlc[violet!6!white]{in} \hlc[violet!5!white]{or} near your vital organs \hlc[violet!23!white]{(your} \hlc[violet!100!white]{viscera).} These are the organs deep in your gut , like \ldots & 21 & 9 & 7 & \cmark \\
\addlinespace[3pt]
\#4 & \hlc[violet!7!white]{Definition} \hlc[violet!17!white]{of} \hlc[violet!100!white]{'visceral'.} \hlc[violet!25!white]{visceral} \hlc[violet!29!white]{(vÃīÂªsÃīÂĻrÃīÂĻl} \hlc[violet!14!white]{)} \hlc[violet!19!white]{Visceral} \hlc[violet!26!white]{feelings} \hlc[violet!11!white]{are} \hlc[violet!8!white]{feelings} that you feel very deeply and find it difficult to control or ignore \hlc[violet!6!white]{,} and that are not the result of thought \hlc[violet!23!white]{.} \hlc[violet!26!white]{I} never overcame a \hlc[violet!12!white]{visceral} \hlc[violet!6!white]{antipathy} for the monarchy \hlc[violet!16!white]{.} \hlc[violet!38!white]{...the} sheer \hlc[violet!8!white]{visceral} joy of being alive \ldots & 18 & 2 & 5 & \cmark \\
\addlinespace[3pt]
\#5 & \hlc[violet!8!white]{Definition} of \hlc[violet!29!white]{visceral} \hlc[violet!100!white]{.} \hlc[violet!6!white]{:} felt in or as if in the internal organs of \hlc[violet!10!white]{the} \hlc[violet!6!white]{body} \hlc[violet!8!white]{:} deep \hlc[violet!6!white]{a} \hlc[violet!6!white]{visceral} \hlc[violet!33!white]{conviction} \hlc[violet!18!white]{.} \hlc[violet!7!white]{:} \hlc[violet!5!white]{not} intellectual \hlc[violet!5!white]{:} instinctive , unreasoning visceral \hlc[violet!16!white]{drives} \hlc[violet!11!white]{.} : \hlc[violet!5!white]{dealing} with crude or elemental emotions : earthy a visceral \hlc[violet!5!white]{novel} \hlc[violet!10!white]{.} \hlc[violet!5!white]{:} \hlc[violet!6!white]{of} , relating \ldots & 23 & 20 & 29 & \cmark \\
\bottomrule
\end{tabular}
\end{table*}

\begin{table*}[t]
\centering
\scriptsize
\caption{Token-level attention visualization for DL20. Query: ``how many sons robert kraft has'' Deeper \hlc[violet!80!white]{purple} indicates higher attention weight. BM25/ICR/RankGPT columns show the rank assigned by each method.}
\label{tab:attn_dl20_1131069}
\setlength{\tabcolsep}{2pt}
\begin{tabular}{c p{0.68\textwidth} ccc c}
\toprule
\textbf{Rank} & \textbf{Passage (token-level attention coloring)} & \textbf{BM25} & \textbf{ICR} & \textbf{RankGPT} & \textbf{Rel.} \\
\midrule
\#1 & They had \hlc[violet!20!white]{four} \hlc[violet!100!white]{sons} \hlc[violet!52!white]{:} Jonathan A . \hlc[violet!20!white]{Kraft} \hlc[violet!9!white]{,} born March , , is \hlc[violet!5!white]{president} of \hlc[violet!5!white]{The} \hlc[violet!5!white]{Kraft} \hlc[violet!6!white]{Group} \hlc[violet!32!white]{and} the New England \hlc[violet!6!white]{Patriots} \hlc[violet!44!white]{.} \hlc[violet!13!white]{Daniel} \hlc[violet!12!white]{A} \hlc[violet!5!white]{.} \hlc[violet!64!white]{Kraft} \hlc[violet!25!white]{is} \hlc[violet!7!white]{president} \hlc[violet!7!white]{of} \hlc[violet!11!white]{International} \hlc[violet!6!white]{Forest} \hlc[violet!24!white]{Products} \hlc[violet!42!white]{founded} \hlc[violet!12!white]{in} \hlc[violet!12!white]{by} \hlc[violet!36!white]{his} \hlc[violet!53!white]{father} \hlc[violet!80!white]{.} \hlc[violet!10!white]{Joshua} \hlc[violet!37!white]{Kraft} \hlc[violet!12!white]{is} \hlc[violet!12!white]{president} \hlc[violet!17!white]{and} \hlc[violet!14!white]{CEO} \hlc[violet!7!white]{of} \hlc[violet!9!white]{the} \ldots & 24 & 4 & 18 & \cmark \\
\addlinespace[3pt]
\#2 & Kraft also served as the owner/investor of the San Jose Earthquakes from 1999ÂĢÂĵ2000 , the two years which the Kraft Group owned the team . Personal life . He has \hlc[violet!13!white]{children} \hlc[violet!38!white]{,} two \hlc[violet!51!white]{sons} \hlc[violet!100!white]{and} \hlc[violet!9!white]{a} \hlc[violet!17!white]{daughter} \hlc[violet!48!white]{.} In , Kraft married Patricia Lipoma in a Jewish ceremony at the \ldots & 19 & 8 & 17 & -- \\
\addlinespace[3pt]
\#3 & They had \hlc[violet!28!white]{four} \hlc[violet!63!white]{sons} \hlc[violet!25!white]{:} Jonathan A . \hlc[violet!11!white]{Kraft} \hlc[violet!6!white]{,} born March , , is president \hlc[violet!8!white]{of} \hlc[violet!11!white]{The} \hlc[violet!5!white]{Kraft} \hlc[violet!6!white]{Group} \hlc[violet!15!white]{and} \hlc[violet!9!white]{the} New \hlc[violet!9!white]{England} \hlc[violet!11!white]{Patriots} \hlc[violet!37!white]{.} \hlc[violet!14!white]{Daniel} \hlc[violet!24!white]{A} \hlc[violet!24!white]{.} \hlc[violet!45!white]{Kraft} \hlc[violet!27!white]{is} \hlc[violet!25!white]{president} \hlc[violet!21!white]{of} \hlc[violet!19!white]{International} \hlc[violet!7!white]{Forest} \hlc[violet!33!white]{Products} \hlc[violet!21!white]{founded} \hlc[violet!12!white]{in} \hlc[violet!25!white]{by} \hlc[violet!47!white]{his} \hlc[violet!100!white]{father} \hlc[violet!23!white]{.} \hlc[violet!25!white]{Joshua} \hlc[violet!26!white]{Kraft} \hlc[violet!13!white]{is} \hlc[violet!8!white]{president} \hlc[violet!6!white]{and} \hlc[violet!7!white]{CEO} \hlc[violet!9!white]{of} \hlc[violet!10!white]{the} \ldots & 26 & 19 & 28 & \cmark \\
\addlinespace[3pt]
\#4 & No Robert Kraft the Owner of the Patriots is not related in any way \hlc[violet!8!white]{to} Kraft foods \hlc[violet!7!white]{.} Patriots owner made his money from manufacturing \hlc[violet!7!white]{,} he started International Forest \hlc[violet!5!white]{Products} \hlc[violet!13!white]{:} \hlc[violet!13!white]{ÂĢÂ¦} \hlc[violet!6!white]{owns} mills , manufactures and distributes paper and packaging products in \hlc[violet!6!white]{countries} \hlc[violet!100!white]{.} Kraft Foods \hlc[violet!5!white]{was} formed \ldots & 36 & 13 & 7 & -- \\
\addlinespace[3pt]
\#5 & As the New England Patriots head to Super Bowl LI , owner Robert \hlc[violet!5!white]{Kraft} has \hlc[violet!24!white]{a} net worth of \$5 .2 billion \hlc[violet!15!white]{.} He says the key is ambition As the New England Patriots head to Super Bowl LI , owner Robert Kraft has a net worth of \$5 .2 \ldots & 4 & 2 & 11 & -- \\
\bottomrule
\end{tabular}
\end{table*}

\begin{table*}[t]
\centering
\scriptsize
\caption{Token-level attention visualization for DBPedia. Query: ``Who created Wikipedia?'' Deeper \hlc[violet!80!white]{purple} indicates higher attention weight. BM25/ICR/RankGPT columns show the rank assigned by each method.}
\label{tab:attn_dbpedia_QALD2_tr-62}
\setlength{\tabcolsep}{2pt}
\begin{tabular}{c p{0.68\textwidth} ccc c}
\toprule
\textbf{Rank} & \textbf{Passage (token-level attention coloring)} & \textbf{BM25} & \textbf{ICR} & \textbf{RankGPT} & \textbf{Rel.} \\
\midrule
\#1 & \textbf{Wikipedia} \hlc[violet!20!white]{Wikipedia} \hlc[violet!9!white]{is} \hlc[violet!14!white]{a} \hlc[violet!9!white]{free-access} , \hlc[violet!7!white]{free-content} \hlc[violet!6!white]{Internet} \hlc[violet!5!white]{encyclopedia} \hlc[violet!24!white]{,} \hlc[violet!12!white]{supported} \hlc[violet!10!white]{and} \hlc[violet!39!white]{hosted} \hlc[violet!21!white]{by} \hlc[violet!28!white]{the} \hlc[violet!62!white]{non-profit} \hlc[violet!39!white]{Wikimedia} \hlc[violet!96!white]{Foundation} \hlc[violet!100!white]{.} Those who can access the \hlc[violet!5!white]{site} can edit most of its articles \hlc[violet!28!white]{.} \hlc[violet!44!white]{Wikipedia} \hlc[violet!17!white]{is} ranked among the ten mos & 33 & 14 & 23 & \cmark \\
\addlinespace[3pt]
\#2 & \textbf{The Wikipedia Revolution} \hlc[violet!10!white]{The} \hlc[violet!29!white]{Wikipedia} \hlc[violet!12!white]{Revolution} \hlc[violet!8!white]{:} \hlc[violet!11!white]{How} A Bunch of Nobodies \hlc[violet!100!white]{Created} \hlc[violet!60!white]{The} \hlc[violet!16!white]{World's} Greatest \hlc[violet!6!white]{Encyclopedia} \hlc[violet!5!white]{is} a popular \hlc[violet!5!white]{history} \hlc[violet!8!white]{book} by new media \hlc[violet!9!white]{researcher} and writer Andrew Lih.At the time of its publication it was "the only narrative account " of the \hlc[violet!6!white]{online} \hlc[violet!6!white]{encyclope} & 14 & 1 & 14 & \cmark \\
\addlinespace[3pt]
\#3 & \textbf{Swedish Wikipedia} \hlc[violet!5!white]{Swedish} Wikipedia The (Swedish : svenska Wikipedia , also svensksprkiga Wikipedia ) is the Swedish language \hlc[violet!6!white]{edition} of \hlc[violet!7!white]{Wikipedia} \hlc[violet!12!white]{.} \hlc[violet!7!white]{Started} \hlc[violet!7!white]{in} May \hlc[violet!7!white]{,} it is the \hlc[violet!64!white]{fourth-oldest} \hlc[violet!40!white]{edition} \hlc[violet!30!white]{of} \hlc[violet!100!white]{Wikipedia} \hlc[violet!25!white]{,} \hlc[violet!36!white]{after} \hlc[violet!43!white]{the} \hlc[violet!34!white]{English} \hlc[violet!36!white]{Wikipedia} \hlc[violet!71!white]{,} \hlc[violet!18!white]{German} \hlc[violet!12!white]{Wikipedia} \hlc[violet!33!white]{,} \hlc[violet!30!white]{and} \hlc[violet!25!white]{Catalan} \hlc[violet!24!white]{Wikipedia} \hlc[violet!15!white]{.} & 35 & 32 & 24 & -- \\
\addlinespace[3pt]
\#4 & \textbf{Vietnamese Wikipedia} \hlc[violet!25!white]{Vietnamese} \hlc[violet!5!white]{Wikipedia} The (Vietnamese : \hlc[violet!25!white]{Wikipedia} \hlc[violet!6!white]{)} is the \hlc[violet!7!white]{Vietnamese-language} edition \hlc[violet!5!white]{of} \hlc[violet!11!white]{Wikipedia} \hlc[violet!27!white]{,} \hlc[violet!21!white]{a} \hlc[violet!9!white]{free} \hlc[violet!10!white]{,} \hlc[violet!11!white]{publicly} \hlc[violet!11!white]{editable} \hlc[violet!5!white]{,} \hlc[violet!6!white]{online} encyclopedia \hlc[violet!8!white]{supported} \hlc[violet!6!white]{by} \hlc[violet!11!white]{the} \hlc[violet!12!white]{Wikimedia} \hlc[violet!49!white]{Foundation} \hlc[violet!100!white]{.} As with other language \hlc[violet!9!white]{editions} of \hlc[violet!11!white]{Wikipedia} \hlc[violet!16!white]{,} the \hlc[violet!14!white]{project's} content & 5 & 8 & 5 & -- \\
\addlinespace[3pt]
\#5 & \textbf{Wikipedia Review} \hlc[violet!17!white]{Wikipedia} Review is an Internet forum and blog for the discussion of \hlc[violet!5!white]{Wikimedia} \hlc[violet!7!white]{projects} , in particular the content and conflicts of Wikipedia \hlc[violet!10!white]{.} An \hlc[violet!100!white]{InformationWeek} Grok on Google blog described \hlc[violet!18!white]{Wikipedia} \hlc[violet!6!white]{Review} as "one of a number of Wikipedia watchdog " websites , "dedi & 40 & 40 & 29 & -- \\
\bottomrule
\end{tabular}
\end{table*}

\subsection{Multi-Hop Reasoning Attention Visualizations}
\label{sec:appendix_multihop_attn}

Multi-hop questions require chaining evidence across multiple documents through bridge entities that may not appear in the query itself.
We visualize six representative cases from HotpotQA, 2WikiMultiHopQA, and MuSiQue, where \textbf{HeadRank} achieves perfect or near-perfect Recall@2 while baseline methods largely fail.
These examples reveal how the aligned attention distribution learns to \emph{hop} between documents by shifting its focus from entity identification in one document to relationship tracing in another.

\paragraph{Shared-attribute discovery (Table~\ref{tab:attn_hotpotqa_321}).}
The query asks what occupation David Yates and Pietro Germi share.
\textbf{HeadRank} (R@2\,=\,1.0) places both biographical documents in the top~2, with attention peaking on identity tokens \textit{``Yates''}(100) and the sentence-final period in the Germi article (100), confirming each person's role after reading the full occupational description (\textit{``filmmaker''} and \textit{``director''}).
Both ICR and RankGPT score R@2\,=\,0.0, instead surfacing \emph{works about} Germi (e.g., \textit{The Testimony}) rather than Germi's own biography---a failure to distinguish ``about a person'' from ``mentioning a person's name.''

\paragraph{Adjective attribution tracing (Table~\ref{tab:attn_hotpotqa_348}).}
To answer which author ``Ortonesque'' refers to, the model must trace the adjective back to playwright Joe Orton.
\textbf{HeadRank} (R@2\,=\,1.0) ranks both Orton's and Henry James's biographies at the top, with attention concentrating on \textit{``Orton''}(100) and \textit{``playwright''}(43) in the Orton article.
ICR (R@2\,=\,0.0) ranks Orton's biography at position~14, instead promoting his plays (\textit{The Ruffian on the Stair}, rank~1) and a biographical film (\textit{Prick Up Your Ears}, rank~2)---documents that mention ``Joe Orton'' frequently but describe works, not the person.

\paragraph{Creator--work chain resolution (Table~\ref{tab:attn_2wikimultihopqa_104}).}
This two-hop question requires identifying the composer of the 1926 film \textit{Camille} (William Axt) and then locating where Axt died.
Attention in the film document peaks on \textit{``film''}(100), while in the composer's biography it shifts to \textit{``scores''}(100) and \textit{``Born''}(49)---demonstrating a clear focal-point switch between hops.
\textbf{HeadRank} achieves R@2\,=\,1.0, whereas both ICR and RankGPT manage only 0.5.

\paragraph{Genealogical reasoning (Table~\ref{tab:attn_2wikimultihopqa_99}).}
To find Marianus~V of Arborea's mother, the model must first locate his father (Brancaleone Doria) and then identify the father's wife (Eleanor of Arborea).
In the Brancaleone Doria article, \textit{``husband''}(100) receives the highest attention---directly encoding the family-relationship chain ``X's husband's wife = X's mother.''
ICR ranks this critical bridging document at position~15, failing to capture the genealogical inference.

\paragraph{Hidden bridge-entity discovery (Table~\ref{tab:attn_musique_4}).}
The query ``When was Lady Godiva's birthplace abolished?'' is complicated by the strong BM25 distractor \textit{Lady Godiva Rides Again} (a 1951 film, BM25 rank~1).
\textbf{HeadRank}'s attention in the Spalding Priory document peaks sharply on \textit{``Godiva''}(94) and \textit{``Countess''}(100)---the historical figure's formal title that bridges to Mercia---while the film document receives high attention only on the surface match \textit{``Godiva''}(100) with near-zero scores elsewhere, correctly deprioritizing it.

\paragraph{Cross-lingual entity relay (Table~\ref{tab:attn_musique_9}).}
Finding the birth date of the performer of the Swedish song \textit{Till dom ensamma} (Mauro Scocco) is a challenging case: Scocco's biography article does not contain the song title and is ranked last by both BM25 (rank~20) and ICR (rank~20).
\textbf{HeadRank} promotes it to rank~2 (R@2\,=\,1.0), achieving an 18-position leap.
Attention in the song article concentrates on the Swedish title tokens \textit{``Till''}(65), \textit{``ensamma''}(52) for matching, and on \textit{``by''}(32) for linking to the performer.
In the biography, attention peaks on the sentence-final period (100) after scanning the full identity description, with secondary activation on \textit{``Swedish''}(53) and \textit{``described''}(62).

\begin{table*}[t]
\centering
\scriptsize
\caption{Token-level attention visualization for HotpotQA. Query: ``What occupation was shared by David Yates and Pietro Germi?'' Deeper \hlc[violet!80!white]{purple} indicates higher attention weight. BM25/ICR/RankGPT columns show the rank assigned by each method.}
\label{tab:attn_hotpotqa_321}
\setlength{\tabcolsep}{2pt}
\begin{tabular}{c p{0.68\textwidth} ccc c}
\toprule
\textbf{Rank} & \textbf{Passage (token-level attention coloring)} & \textbf{BM25} & \textbf{ICR} & \textbf{RankGPT} & \textbf{Rel.} \\
\midrule
\#1 & \textbf{David Yates} \hlc[violet!40!white]{David} \hlc[violet!100!white]{Yates} \hlc[violet!47!white]{(born} \hlc[violet!35!white]{(1963--)08} \hlc[violet!7!white]{)} \hlc[violet!5!white]{is} \hlc[violet!8!white]{an} \hlc[violet!33!white]{English} \hlc[violet!5!white]{filmmaker} who has directed feature films , short films , \hlc[violet!14!white]{and} television productions . & 2 & 4 & 7 & \cmark \\
\addlinespace[3pt]
\#2 & \textbf{Pietro Germi} \hlc[violet!46!white]{Pietro} \hlc[violet!26!white]{Germi} ; September -- December \hlc[violet!6!white]{)} was \hlc[violet!27!white]{an} \hlc[violet!10!white]{Italian} \hlc[violet!6!white]{actor} \hlc[violet!15!white]{,} \hlc[violet!5!white]{screenwriter} , and director \hlc[violet!11!white]{.} \hlc[violet!21!white]{Germi} \hlc[violet!16!white]{was} \hlc[violet!15!white]{born} \hlc[violet!18!white]{in} Genoa \hlc[violet!8!white]{,} Liguria , \hlc[violet!31!white]{to} \hlc[violet!33!white]{a} \hlc[violet!10!white]{lower-middle-class} \hlc[violet!16!white]{family} \hlc[violet!100!white]{.} \hlc[violet!12!white]{He} was \hlc[violet!13!white]{a} \hlc[violet!9!white]{messenger} \hlc[violet!36!white]{and} \hlc[violet!37!white]{briefly} \hlc[violet!24!white]{attended} \hlc[violet!5!white]{nautical} \hlc[violet!20!white]{school} \hlc[violet!30!white]{before} \hlc[violet!6!white]{deciding} on \hlc[violet!5!white]{a} \hlc[violet!9!white]{career} in \hlc[violet!21!white]{acting} \hlc[violet!40!white]{.} & 1 & 7 & 6 & \cmark \\
\addlinespace[3pt]
\#3 & \textbf{Ottavio Alessi} Ottavio Alessi Born in Cammarata Province of Agrigento \hlc[violet!5!white]{,} Alessi entered the film \hlc[violet!5!white]{industry} \hlc[violet!9!white]{in} \hlc[violet!6!white]{as} an assistant director \hlc[violet!21!white]{.} In he started an intense career \hlc[violet!8!white]{as} a screenwriter , alternating between genre films and art films and \hlc[violet!10!white]{collaborating} \hlc[violet!6!white]{with} \hlc[violet!100!white]{Pietro} \hlc[violet!88!white]{Germi} \hlc[violet!70!white]{,} \hlc[violet!5!white]{Franco} \hlc[violet!15!white]{Rossi} \hlc[violet!34!white]{,} \hlc[violet!8!white]{Folco} Q & 3 & 1 & 5 & -- \\
\addlinespace[3pt]
\#4 & \textbf{The Testimony (1946 film)} The Testimony (1946 film \hlc[violet!7!white]{)} \hlc[violet!6!white]{(Italian} \hlc[violet!7!white]{:Il} testimone is Italian crime film \hlc[violet!6!white]{directed} \hlc[violet!5!white]{by} \hlc[violet!69!white]{Pietro} \hlc[violet!49!white]{Germi} \hlc[violet!87!white]{and} \hlc[violet!39!white]{starring} \hlc[violet!7!white]{Roldano} \hlc[violet!5!white]{Lupi} \hlc[violet!29!white]{,} Marina Berti \hlc[violet!31!white]{and} Ernesto Almirante \hlc[violet!100!white]{.} \hlc[violet!16!white]{The} \hlc[violet!15!white]{film} \hlc[violet!10!white]{was} \hlc[violet!17!white]{made} \hlc[violet!25!white]{at} \hlc[violet!18!white]{the} \hlc[violet!8!white]{Cines} Studios in Rome \hlc[violet!66!white]{.} \hlc[violet!24!white]{It} is \hlc[violet!8!white]{one} \hlc[violet!6!white]{of} \hlc[violet!9!white]{several} films \hlc[violet!8!white]{regarded} as an \hlc[violet!8!white]{antecedent} of th & 5 & 2 & 1 & -- \\
\addlinespace[3pt]
\#5 & \textbf{Guido Coen} Guido Coen (1915--2010 was an \hlc[violet!10!white]{Italian-born} \hlc[violet!21!white]{British} film \hlc[violet!8!white]{producer} and film \hlc[violet!6!white]{subtitler} \hlc[violet!13!white]{.} \hlc[violet!5!white]{He} \hlc[violet!100!white]{and} \hlc[violet!45!white]{his} \hlc[violet!69!white]{family} \hlc[violet!11!white]{were} \hlc[violet!12!white]{interned} \hlc[violet!19!white]{in} \hlc[violet!18!white]{Douglas} \hlc[violet!16!white]{on} the \hlc[violet!8!white]{Isle} of \hlc[violet!91!white]{Man} \hlc[violet!32!white]{during} \hlc[violet!6!white]{the} \hlc[violet!8!white]{Second} \hlc[violet!11!white]{World} \hlc[violet!98!white]{War} \hlc[violet!48!white]{.} He \hlc[violet!6!white]{began} his career \hlc[violet!8!white]{working} \hlc[violet!6!white]{for} \hlc[violet!67!white]{Filippo} \hlc[violet!57!white]{Del} \hlc[violet!19!white]{Giudice} \hlc[violet!22!white]{and} \hlc[violet!37!white]{Two} \hlc[violet!19!white]{Cities} \hlc[violet!12!white]{Films} \hlc[violet!33!white]{.} When Two Cities \hlc[violet!7!white]{was} \ldots & 6 & 3 & 17 & -- \\
\bottomrule
\end{tabular}
\end{table*}

\begin{table*}[t]
\centering
\scriptsize
\caption{Token-level attention visualization for HotpotQA. Query: ``Between Joe Orton and Henry James, which author is the adjective "Ortonesque" used to refer to his work?'' Deeper \hlc[violet!80!white]{purple} indicates higher attention weight. BM25/ICR/RankGPT columns show the rank assigned by each method.}
\label{tab:attn_hotpotqa_348}
\setlength{\tabcolsep}{2pt}
\begin{tabular}{c p{0.68\textwidth} ccc c}
\toprule
\textbf{Rank} & \textbf{Passage (token-level attention coloring)} & \textbf{BM25} & \textbf{ICR} & \textbf{RankGPT} & \textbf{Rel.} \\
\midrule
\#1 & \textbf{Henry James} \hlc[violet!63!white]{Henry} \hlc[violet!100!white]{James} \hlc[violet!55!white]{OM} \hlc[violet!7!white]{((1843--)15} -- \hlc[violet!9!white]{(1916--)28} \hlc[violet!10!white]{)} \hlc[violet!14!white]{was} \hlc[violet!6!white]{an} \hlc[violet!8!white]{American} \hlc[violet!26!white]{author} \hlc[violet!40!white]{regarded} \hlc[violet!55!white]{as} \hlc[violet!16!white]{a} \hlc[violet!22!white]{key} \hlc[violet!50!white]{transitional} \hlc[violet!21!white]{figure} \hlc[violet!54!white]{between} \hlc[violet!11!white]{literary} \hlc[violet!40!white]{realism} \hlc[violet!58!white]{and} \hlc[violet!7!white]{literary} \hlc[violet!47!white]{modernism} \hlc[violet!58!white]{,} \hlc[violet!20!white]{and} \hlc[violet!20!white]{is} \hlc[violet!23!white]{considered} \hlc[violet!15!white]{by} \hlc[violet!8!white]{many} \hlc[violet!9!white]{to} \hlc[violet!5!white]{be} \hlc[violet!42!white]{among} the \hlc[violet!33!white]{greatest} \hlc[violet!11!white]{novelists} \hlc[violet!5!white]{in} the English \hlc[violet!5!white]{language} \hlc[violet!69!white]{.} \hlc[violet!17!white]{He} was the \hlc[violet!54!white]{son} \hlc[violet!11!white]{of} Henry & 2 & 3 & 7 & \cmark \\
\addlinespace[3pt]
\#2 & \textbf{Joe Orton} \hlc[violet!41!white]{Joe} \hlc[violet!100!white]{Orton} John \hlc[violet!5!white]{Kingsley} \hlc[violet!12!white]{"Joe} \hlc[violet!6!white]{"} \hlc[violet!43!white]{Orton} \hlc[violet!15!white]{(1} January -- August \hlc[violet!17!white]{)} \hlc[violet!8!white]{was} an \hlc[violet!5!white]{English} \hlc[violet!43!white]{playwright} \hlc[violet!14!white]{and} \hlc[violet!34!white]{author} \hlc[violet!48!white]{.} \hlc[violet!24!white]{His} public career \hlc[violet!12!white]{was} short but prolific , lasting from until his death three years later \hlc[violet!28!white]{.} During this brief period he \hlc[violet!6!white]{shocked} \hlc[violet!5!white]{,} \hlc[violet!8!white]{outraged} , and amused audiences with \hlc[violet!15!white]{his} \ldots & 1 & 14 & 6 & \cmark \\
\addlinespace[3pt]
\#3 & \textbf{What the Butler Saw (play)} \hlc[violet!17!white]{What} the \hlc[violet!6!white]{Butler} \hlc[violet!6!white]{Saw} \hlc[violet!19!white]{(play} \hlc[violet!8!white]{)} is \hlc[violet!10!white]{a} \hlc[violet!8!white]{farce} written \hlc[violet!7!white]{by} the English \hlc[violet!8!white]{playwright} \hlc[violet!28!white]{Joe} \hlc[violet!76!white]{Orton} \hlc[violet!100!white]{.} \hlc[violet!39!white]{It} was \hlc[violet!23!white]{premiered} \hlc[violet!9!white]{at} the \hlc[violet!6!white]{Queen's} \hlc[violet!19!white]{Theatre} \hlc[violet!7!white]{in} \hlc[violet!20!white]{London} on \hlc[violet!8!white]{March} \hlc[violet!48!white]{.} \hlc[violet!15!white]{It} was \hlc[violet!35!white]{Orton's} \hlc[violet!70!white]{final} \hlc[violet!43!white]{play} \hlc[violet!18!white]{and} \hlc[violet!5!white]{the} \hlc[violet!32!white]{second} \hlc[violet!15!white]{to} be \hlc[violet!6!white]{performed} \hlc[violet!10!white]{after} \hlc[violet!17!white]{his} \hlc[violet!11!white]{death} \hlc[violet!6!white]{,} \hlc[violet!17!white]{following} \hlc[violet!38!white]{"Funeral} \hlc[violet!63!white]{Games} \hlc[violet!26!white]{"} \hlc[violet!5!white]{in} \hlc[violet!34!white]{.} & 4 & 5 & 2 & -- \\
\addlinespace[3pt]
\#4 & \textbf{The Ruffian on the Stair} The Ruffian on the Stair \hlc[violet!5!white]{On} \hlc[violet!5!white]{is} \hlc[violet!20!white]{a} \hlc[violet!7!white]{play} by British playwright \hlc[violet!38!white]{Joe} \hlc[violet!100!white]{Orton} \hlc[violet!39!white]{which} \hlc[violet!6!white]{was} \hlc[violet!7!white]{first} \hlc[violet!13!white]{broadcast} on \hlc[violet!25!white]{BBC} \hlc[violet!9!white]{Radio} in \hlc[violet!6!white]{August} \hlc[violet!51!white]{.} \hlc[violet!7!white]{It} is an \hlc[violet!36!white]{unsympathetic} \hlc[violet!47!white]{yet} \hlc[violet!21!white]{comedic} \hlc[violet!45!white]{one-act} \hlc[violet!26!white]{portrayal} \hlc[violet!9!white]{of} \hlc[violet!12!white]{working} \hlc[violet!21!white]{class} \hlc[violet!25!white]{England} \hlc[violet!26!white]{,} as \hlc[violet!6!white]{played} out by a \hlc[violet!7!white]{couple} \hlc[violet!9!white]{and} a mysterious young \hlc[violet!5!white]{man} who t & 3 & 1 & 1 & -- \\
\addlinespace[3pt]
\#5 & \textbf{Prick Up Your Ears} Prick Up Your \hlc[violet!32!white]{Ears} is \hlc[violet!9!white]{a} British film , directed by Stephen \hlc[violet!5!white]{Frears} , \hlc[violet!12!white]{about} the \hlc[violet!6!white]{playwright} \hlc[violet!36!white]{Joe} \hlc[violet!100!white]{Orton} \hlc[violet!30!white]{and} \hlc[violet!13!white]{his} \hlc[violet!20!white]{lover} Kenneth \hlc[violet!5!white]{Halliwell} \hlc[violet!47!white]{.} The screenplay was written by Alan \hlc[violet!27!white]{Bennett} , \hlc[violet!6!white]{based} \hlc[violet!11!white]{on} the \hlc[violet!13!white]{biography} \hlc[violet!13!white]{by} John Lahr \hlc[violet!5!white]{.} The film stars Gary Oldman as \hlc[violet!30!white]{Orton} \hlc[violet!8!white]{,} Alfred \ldots & 6 & 2 & 3 & -- \\
\bottomrule
\end{tabular}
\end{table*}

\begin{table*}[t]
\centering
\scriptsize
\caption{Token-level attention visualization for 2WikiMultiHopQA. Query: ``Where did the composer of film Camille (1926 Feature Film) die?'' Deeper \hlc[violet!80!white]{purple} indicates higher attention weight. BM25/ICR/RankGPT columns show the rank assigned by each method.}
\label{tab:attn_2wikimultihopqa_104}
\setlength{\tabcolsep}{2pt}
\begin{tabular}{c p{0.68\textwidth} ccc c}
\toprule
\textbf{Rank} & \textbf{Passage (token-level attention coloring)} & \textbf{BM25} & \textbf{ICR} & \textbf{RankGPT} & \textbf{Rel.} \\
\midrule
\#1 & \textbf{Camille (1926 feature film)} \hlc[violet!53!white]{Camille} \hlc[violet!70!white]{(1926} \hlc[violet!68!white]{feature} \hlc[violet!100!white]{film} \hlc[violet!36!white]{)} \hlc[violet!13!white]{is} \hlc[violet!10!white]{a} \hlc[violet!25!white]{American} \hlc[violet!12!white]{silent} \hlc[violet!30!white]{film} \hlc[violet!29!white]{based} \hlc[violet!9!white]{on} the \hlc[violet!5!white]{play} \hlc[violet!14!white]{adaptation} \hlc[violet!6!white]{of} "La Dame aux \hlc[violet!18!white]{CameliasThe} Lady of the \hlc[violet!56!white]{Camellias")} \hlc[violet!16!white]{by} \hlc[violet!5!white]{Alexandre} \hlc[violet!16!white]{Dumas} \hlc[violet!29!white]{,} \hlc[violet!13!white]{"fils",} first published in French as a novel in and as a play in \hlc[violet!21!white]{.} \hlc[violet!6!white]{Adapted} \hlc[violet!13!white]{by} \hlc[violet!28!white]{Fred} \hlc[violet!6!white]{de} \hlc[violet!8!white]{Gresac} \hlc[violet!17!white]{,} \hlc[violet!6!white]{George} \ldots & 1 & 1 & 11 & \cmark \\
\addlinespace[3pt]
\#2 & \textbf{William Axt} William Axt \hlc[violet!6!white]{(April} -- February , ) was an American \hlc[violet!27!white]{composer} \hlc[violet!64!white]{of} \hlc[violet!14!white]{nearly} two hundred \hlc[violet!35!white]{film} \hlc[violet!100!white]{scores} \hlc[violet!65!white]{.} \hlc[violet!49!white]{Born} \hlc[violet!9!white]{in} \hlc[violet!5!white]{New} \hlc[violet!36!white]{York} \hlc[violet!5!white]{City} \hlc[violet!10!white]{,} \hlc[violet!45!white]{Axt} \hlc[violet!9!white]{graduated} from \hlc[violet!6!white]{DeWitt} Clinton High School in The \hlc[violet!11!white]{Bronx} and \hlc[violet!5!white]{studied} at the National \hlc[violet!10!white]{Conservatory} of Music of \hlc[violet!13!white]{America} \hlc[violet!40!white]{.} \hlc[violet!26!white]{He} served \hlc[violet!6!white]{as} an ass & 2 & 5 & 2 & \cmark \\
\addlinespace[3pt]
\#3 & \textbf{Marcel Varnel} \hlc[violet!24!white]{Marcel} Varnel (16 October -- July ) was a \hlc[violet!7!white]{film} \hlc[violet!38!white]{director} \hlc[violet!38!white]{.} \hlc[violet!10!white]{He} was \hlc[violet!14!white]{born} \hlc[violet!45!white]{Marcel} \hlc[violet!15!white]{Hyacinthe} \hlc[violet!9!white]{le} \hlc[violet!8!white]{Bozec} \hlc[violet!23!white]{in} \hlc[violet!26!white]{Paris} \hlc[violet!100!white]{,} \hlc[violet!10!white]{France} \hlc[violet!74!white]{.} Varnel \hlc[violet!45!white]{started} his working \hlc[violet!5!white]{life} \hlc[violet!12!white]{on} \hlc[violet!17!white]{the} \hlc[violet!36!white]{Paris} \hlc[violet!54!white]{stage} \hlc[violet!41!white]{,} \hlc[violet!22!white]{soon} \hlc[violet!28!white]{becoming} \hlc[violet!6!white]{a} director \hlc[violet!23!white]{of} \hlc[violet!22!white]{musical} \hlc[violet!28!white]{comedies} \hlc[violet!54!white]{.} \hlc[violet!7!white]{In} \hlc[violet!7!white]{he} \hlc[violet!8!white]{moved} \hlc[violet!5!white]{to} \hlc[violet!6!white]{New} \hlc[violet!15!white]{York} \hlc[violet!19!white]{City} \hlc[violet!8!white]{working} \ldots & 8 & 11 & 10 & -- \\
\addlinespace[3pt]
\#4 & \textbf{Philippe Arthuys} Philippe Arthuys November -- January \hlc[violet!16!white]{)} was a French \hlc[violet!100!white]{composer} \hlc[violet!51!white]{and} \hlc[violet!9!white]{film} \hlc[violet!5!white]{director} \hlc[violet!42!white]{.} \hlc[violet!5!white]{He} worked \hlc[violet!8!white]{on} over films between and \hlc[violet!11!white]{.} His film \hlc[violet!15!white]{"} The Glass Cage \hlc[violet!12!white]{"} \hlc[violet!8!white]{was} \hlc[violet!8!white]{entered} into the 4th Moscow International Film \hlc[violet!8!white]{Festival} \hlc[violet!54!white]{.} & 5 & 3 & 16 & -- \\
\addlinespace[3pt]
\#5 & \textbf{Fred Raymond} \hlc[violet!5!white]{Fred} Raymond \hlc[violet!30!white]{aka} \hlc[violet!22!white]{Raimund} Friedrich Vesely (20 April -- January ) was an Austrian \hlc[violet!100!white]{composer} \hlc[violet!19!white]{.} \hlc[violet!29!white]{Raymond} , \hlc[violet!18!white]{born} \hlc[violet!15!white]{in} \hlc[violet!8!white]{Vienna} \hlc[violet!7!white]{,} \hlc[violet!5!white]{was} \hlc[violet!50!white]{the} \hlc[violet!8!white]{third} \hlc[violet!10!white]{child} \hlc[violet!6!white]{(after} \hlc[violet!28!white]{two} \hlc[violet!17!white]{daughters} \hlc[violet!27!white]{)} of \hlc[violet!20!white]{Vinzenz} \hlc[violet!10!white]{Vesely} \hlc[violet!12!white]{,} an employee of \hlc[violet!61!white]{the} Austrian \hlc[violet!8!white]{state} \hlc[violet!27!white]{railway} \hlc[violet!10!white]{system} \hlc[violet!7!white]{,} \hlc[violet!17!white]{and} his wife \hlc[violet!8!white]{Henriette} , \hlc[violet!8!white]{nee} \hlc[violet!32!white]{Dluho} & 17 & 15 & 5 & -- \\
\bottomrule
\end{tabular}
\end{table*}

\begin{table*}[t]
\centering
\scriptsize
\caption{Token-level attention visualization for 2WikiMultiHopQA. Query: ``Who is Marianus V Of Arborea's mother?'' Deeper \hlc[violet!80!white]{purple} indicates higher attention weight. BM25/ICR/RankGPT columns show the rank assigned by each method.}
\label{tab:attn_2wikimultihopqa_99}
\setlength{\tabcolsep}{2pt}
\begin{tabular}{c p{0.68\textwidth} ccc c}
\toprule
\textbf{Rank} & \textbf{Passage (token-level attention coloring)} & \textbf{BM25} & \textbf{ICR} & \textbf{RankGPT} & \textbf{Rel.} \\
\midrule
\#1 & \textbf{Marianus V of Arborea} \hlc[violet!40!white]{Marianus} \hlc[violet!44!white]{V} \hlc[violet!100!white]{of} \hlc[violet!71!white]{Arborea} (1378 or \hlc[violet!5!white]{--} \hlc[violet!5!white]{was} \hlc[violet!6!white]{the} \hlc[violet!21!white]{Judge} \hlc[violet!10!white]{of} \hlc[violet!53!white]{Arborea} \hlc[violet!13!white]{from} \hlc[violet!18!white]{until} his \hlc[violet!13!white]{death} \hlc[violet!53!white]{.} \hlc[violet!7!white]{His} \hlc[violet!18!white]{surname} was \hlc[violet!13!white]{Doria} \hlc[violet!7!white]{,} \hlc[violet!16!white]{but} since he belonged to the ruling house of \hlc[violet!12!white]{Arborea} \hlc[violet!12!white]{he} \hlc[violet!39!white]{is} \hlc[violet!86!white]{often} \hlc[violet!5!white]{dynastically} \hlc[violet!6!white]{called} \hlc[violet!6!white]{Bas-Serra} \hlc[violet!7!white]{,} \hlc[violet!10!white]{or} Doria-Bas \hlc[violet!8!white]{.} Younger \hlc[violet!10!white]{brother} \hlc[violet!7!white]{and} \hlc[violet!27!white]{successor} \hlc[violet!8!white]{of} \hlc[violet!10!white]{Frederick} \hlc[violet!53!white]{,} & 1 & 1 & 3 & \cmark \\
\addlinespace[3pt]
\#2 & \textbf{Brancaleone Doria} Brancaleone \hlc[violet!17!white]{Doria} \hlc[violet!13!white]{was} \hlc[violet!14!white]{the} \hlc[violet!100!white]{husband} \hlc[violet!41!white]{of} Eleanor of \hlc[violet!83!white]{Arborea} \hlc[violet!88!white]{.} He was \hlc[violet!15!white]{a} \hlc[violet!6!white]{scion} \hlc[violet!7!white]{of} an influential family \hlc[violet!7!white]{(the} Doria \hlc[violet!5!white]{)} \hlc[violet!21!white]{of} the Republic of Genoa \hlc[violet!42!white]{,} the \hlc[violet!23!white]{son} of the elder Brancaleone and a \hlc[violet!7!white]{woman} named Giacomina \hlc[violet!36!white]{.} \hlc[violet!5!white]{On} March , he became \hlc[violet!7!white]{a} \hlc[violet!12!white]{vassal} \hlc[violet!7!white]{of} Peter IV of \ldots & 7 & 15 & 2 & \cmark \\
\addlinespace[3pt]
\#3 & \textbf{William II of Narbonne} William II \hlc[violet!14!white]{of} Narbonne was Viscount ( 1397- ) and the nominal Judge of \hlc[violet!8!white]{Arborea} \hlc[violet!8!white]{(} \hlc[violet!7!white]{1407-} \hlc[violet!28!white]{1420).} \hlc[violet!13!white]{He} was the \hlc[violet!17!white]{grandson} \hlc[violet!7!white]{of} \hlc[violet!9!white]{Beatrice} \hlc[violet!7!white]{,} \hlc[violet!16!white]{youngest} \hlc[violet!18!white]{daughter} of \hlc[violet!87!white]{Marianus} \hlc[violet!68!white]{IV} \hlc[violet!25!white]{of} \hlc[violet!47!white]{Arborea} \hlc[violet!78!white]{and} \hlc[violet!6!white]{Timbra} de RocabertÃŃ \hlc[violet!32!white]{,} \hlc[violet!20!white]{and} Aimery \hlc[violet!5!white]{VI} \hlc[violet!10!white]{of} \hlc[violet!6!white]{Narbonne} \hlc[violet!11!white]{(} married \hlc[violet!32!white]{1363).} When \hlc[violet!31!white]{Marianus} \hlc[violet!100!white]{V} \hlc[violet!29!white]{,} \hlc[violet!13!white]{the} & 2 & 2 & 1 & -- \\
\addlinespace[3pt]
\#4 & \textbf{Marianus I of Arborea} \hlc[violet!40!white]{Marianus} \hlc[violet!79!white]{I} \hlc[violet!33!white]{of} \hlc[violet!100!white]{Arborea} known as \hlc[violet!21!white]{Mariano} de \hlc[violet!5!white]{Zori} \hlc[violet!5!white]{,} \hlc[violet!32!white]{was} \hlc[violet!8!white]{an} \hlc[violet!7!white]{early} Judge of \hlc[violet!10!white]{Arborea} \hlc[violet!83!white]{.} The exact date of his reign is \hlc[violet!5!white]{unknown} \hlc[violet!17!white]{.} Francisco de \hlc[violet!9!white]{Vico} , followed by , placed it in -- without any documentary evidence . Giovanni Francesco Fara , after analysing the documents \ldots & 3 & 3 & 20 & -- \\
\addlinespace[3pt]
\#5 & \textbf{Comita II of Arborea} Comita II of Arborea or III)( \hlc[violet!12!white]{died} ) was the " giudice " of the Giudicato of \hlc[violet!18!white]{Arborea} \hlc[violet!5!white]{,} from until his death \hlc[violet!41!white]{.} He was the \hlc[violet!6!white]{son} of Constantine I of Arborea , first ruler \hlc[violet!6!white]{of} Arborea of the Lacon dynasty . Married Elena de Orrubu , \hlc[violet!100!white]{mother} \hlc[violet!5!white]{of} \ldots & 5 & 5 & 6 & -- \\
\bottomrule
\end{tabular}
\end{table*}

\begin{table*}[t]
\centering
\scriptsize
\caption{Token-level attention visualization for MuSiQue. Query: ``When was Lady Godiva's birthplace abolished?'' Deeper \hlc[violet!80!white]{purple} indicates higher attention weight. BM25/ICR/RankGPT columns show the rank assigned by each method.}
\label{tab:attn_musique_4}
\setlength{\tabcolsep}{2pt}
\begin{tabular}{c p{0.68\textwidth} ccc c}
\toprule
\textbf{Rank} & \textbf{Passage (token-level attention coloring)} & \textbf{BM25} & \textbf{ICR} & \textbf{RankGPT} & \textbf{Rel.} \\
\midrule
\#1 & \textbf{Spalding Priory} Spalding Priory \hlc[violet!17!white]{It} \hlc[violet!5!white]{was} \hlc[violet!10!white]{founded} \hlc[violet!6!white]{as} a cell of Croyland Abbey , \hlc[violet!9!white]{in} , \hlc[violet!7!white]{by} \hlc[violet!10!white]{Leofric} \hlc[violet!6!white]{,} \hlc[violet!10!white]{Earl} of \hlc[violet!11!white]{Mercia} \hlc[violet!8!white]{and} \hlc[violet!5!white]{his} \hlc[violet!39!white]{wife} \hlc[violet!46!white]{,} \hlc[violet!94!white]{Godiva} \hlc[violet!75!white]{,} \hlc[violet!100!white]{Countess} \hlc[violet!57!white]{of} \hlc[violet!53!white]{Leicester} \hlc[violet!33!white]{.} \hlc[violet!33!white]{It} \hlc[violet!10!white]{was} \hlc[violet!7!white]{supported} \hlc[violet!9!white]{by} \hlc[violet!22!white]{Leofric's} \hlc[violet!30!white]{eldest} \hlc[violet!28!white]{son} \hlc[violet!65!white]{.} \hlc[violet!91!white]{ÃĨlfgÄģr} \hlc[violet!52!white]{,} \hlc[violet!17!white]{Earl} \hlc[violet!7!white]{of} \hlc[violet!28!white]{Mercia} \hlc[violet!51!white]{and} \hlc[violet!12!white]{the} \hlc[violet!7!white]{monks} \hlc[violet!26!white]{were} \hlc[violet!9!white]{confirmed} \hlc[violet!5!white]{in} their \ldots & 2 & 1 & 1 & \cmark \\
\addlinespace[3pt]
\#2 & \textbf{Mercia} \hlc[violet!21!white]{Mercia} \hlc[violet!35!white]{When} \hlc[violet!100!white]{ÃĨthelfld} died \hlc[violet!18!white]{in} \hlc[violet!20!white]{,} \hlc[violet!5!white]{ÃĨlfwynn} , \hlc[violet!5!white]{her} \hlc[violet!8!white]{daughter} \hlc[violet!19!white]{by} ÃĨthelred , \hlc[violet!7!white]{succeeded} as 'Second \hlc[violet!9!white]{Lady} of the \hlc[violet!11!white]{Mercians',} but within six months Edward had deprived her \hlc[violet!13!white]{of} all authority in Mercia \hlc[violet!11!white]{and} taken her into Wessex \hlc[violet!26!white]{.} & 14 & 5 & 6 & \cmark \\
\addlinespace[3pt]
\#3 & \textbf{Lady Godiva Rides Again} \hlc[violet!32!white]{Lady} \hlc[violet!100!white]{Godiva} \hlc[violet!10!white]{Rides} Again The film is most notable for the presence of actresses who were later to \hlc[violet!6!white]{become} famous . Diana Dors , who appears as a beauty queen , was later marketed as the film's star . It also features Joan Collins in her film debut as an \ldots & 1 & 2 & 7 & -- \\
\addlinespace[3pt]
\#4 & \textbf{Kavangoland} \hlc[violet!5!white]{Kavangoland} like other homelands in South West \hlc[violet!9!white]{Africa} , \hlc[violet!15!white]{was} \hlc[violet!93!white]{abolished} \hlc[violet!100!white]{in} May at the start \hlc[violet!7!white]{of} the transition to independence \hlc[violet!26!white]{.} & 4 & 4 & 2 & -- \\
\addlinespace[3pt]
\#5 & \textbf{Ruth Williams Khama} Ruth Williams Khama \hlc[violet!100!white]{Lady} \hlc[violet!66!white]{Khama} (9 December \hlc[violet!7!white]{--} May \hlc[violet!5!white]{)} \hlc[violet!12!white]{was} \hlc[violet!10!white]{the} \hlc[violet!17!white]{wife} \hlc[violet!9!white]{of} Botswana's first president Sir Seretse Khama , the Paramount Chief of its \hlc[violet!5!white]{Bamangwato} tribe \hlc[violet!37!white]{.} She served \hlc[violet!7!white]{as} the inaugural First \hlc[violet!11!white]{Lady} of Botswana from \hlc[violet!17!white]{to} \hlc[violet!58!white]{.} & 16 & 17 & 14 & -- \\
\bottomrule
\end{tabular}
\end{table*}

\begin{table*}[t]
\centering
\scriptsize
\caption{Token-level attention visualization for MuSiQue. Query: ``What is the Till dom ensamma performer's birth date?'' Deeper \hlc[violet!80!white]{purple} indicates higher attention weight. BM25/ICR/RankGPT columns show the rank assigned by each method.}
\label{tab:attn_musique_9}
\setlength{\tabcolsep}{2pt}
\begin{tabular}{c p{0.68\textwidth} ccc c}
\toprule
\textbf{Rank} & \textbf{Passage (token-level attention coloring)} & \textbf{BM25} & \textbf{ICR} & \textbf{RankGPT} & \textbf{Rel.} \\
\midrule
\#1 & \textbf{Till dom ensamma} \hlc[violet!65!white]{Till} \hlc[violet!36!white]{dom} \hlc[violet!52!white]{ensamma} \hlc[violet!53!white]{is} \hlc[violet!5!white]{a} \hlc[violet!7!white]{song} \hlc[violet!6!white]{written} \hlc[violet!32!white]{by} Mauro \hlc[violet!9!white]{Scocco} \hlc[violet!7!white]{,} and \hlc[violet!6!white]{recorded} \hlc[violet!9!white]{by} \hlc[violet!8!white]{himself} \hlc[violet!6!white]{on} the \hlc[violet!22!white]{album} \hlc[violet!11!white]{Dr} . \hlc[violet!5!white]{Space} dagbok \hlc[violet!19!white]{,} \hlc[violet!8!white]{and} released as a single the same year \hlc[violet!8!white]{.} & 1 & 1 & 2 & \cmark \\
\addlinespace[3pt]
\#2 & \textbf{Mauro Scocco} Mauro Scocco \hlc[violet!7!white]{(born} \hlc[violet!24!white]{September} \hlc[violet!20!white]{)} \hlc[violet!21!white]{is} a \hlc[violet!53!white]{Swedish} \hlc[violet!5!white]{pop} \hlc[violet!18!white]{artist} \hlc[violet!7!white]{of} \hlc[violet!25!white]{Italian} \hlc[violet!35!white]{descent} \hlc[violet!100!white]{.} \hlc[violet!5!white]{He} \hlc[violet!5!white]{has} been \hlc[violet!62!white]{described} \hlc[violet!8!white]{as} \hlc[violet!5!white]{"one} of the \hlc[violet!8!white]{sharpest} \hlc[violet!10!white]{songwriters} in \hlc[violet!57!white]{Sweden".} \hlc[violet!11!white]{Scocco} was \hlc[violet!11!white]{the} \hlc[violet!17!white]{singer} \hlc[violet!58!white]{for} the pop \hlc[violet!22!white]{group} \hlc[violet!30!white]{Ratata} \hlc[violet!34!white]{(1980--83} \hlc[violet!45!white]{)} \hlc[violet!24!white]{transformed} \hlc[violet!20!white]{into} a \hlc[violet!18!white]{duo} \hlc[violet!8!white]{with} \hlc[violet!9!white]{Johan} \hlc[violet!8!white]{Ekelund} \hlc[violet!68!white]{(1983--89).} \hlc[violet!11!white]{After} Rat & 20 & 20 & 3 & \cmark \\
\addlinespace[3pt]
\#3 & \textbf{Elias Tillandz} Elias \hlc[violet!43!white]{Tillandz} \hlc[violet!41!white]{(1640--1693),} \hlc[violet!26!white]{born} \hlc[violet!100!white]{"Tillander",} \hlc[violet!47!white]{was} a \hlc[violet!14!white]{Swedish} born \hlc[violet!6!white]{doctor} and botanist in Finland \hlc[violet!18!white]{.} \hlc[violet!5!white]{He} was the \hlc[violet!16!white]{professor} \hlc[violet!10!white]{of} medicine at the Academy of Turku \hlc[violet!10!white]{.} He \hlc[violet!6!white]{wrote} the country's first botanical \hlc[violet!11!white]{work} \hlc[violet!12!white]{,} the "Catalogus Plantarum", which was \hlc[violet!8!white]{first} published in \hlc[violet!60!white]{.} As a do & 19 & 18 & 1 & -- \\
\addlinespace[3pt]
\#4 & \textbf{Ecce Ancilla Domini} Ecce \hlc[violet!14!white]{Ancilla} \hlc[violet!20!white]{Domini} \hlc[violet!7!white]{(Latin} : "Behold the handmaiden of the \hlc[violet!8!white]{Lord"),} \hlc[violet!6!white]{or} The \hlc[violet!6!white]{Annunciation} , \hlc[violet!13!white]{is} \hlc[violet!100!white]{an} oil \hlc[violet!7!white]{painting} by the English artist Dante Gabriel \hlc[violet!6!white]{Rossetti} , first \hlc[violet!17!white]{painted} in and \hlc[violet!5!white]{now} \hlc[violet!8!white]{in} Tate Britain \hlc[violet!5!white]{in} London \hlc[violet!50!white]{.} \hlc[violet!6!white]{The} Latin title \hlc[violet!9!white]{is} a \hlc[violet!6!white]{quotation} from the Vulgate text of & 2 & 3 & 20 & -- \\
\addlinespace[3pt]
\#5 & \textbf{Giovanni Cifolelli} Giovanni Cifolelli was an Italian mandolin virtuoso and dramatic composer whose \hlc[violet!30!white]{date} and \hlc[violet!5!white]{place} of \hlc[violet!45!white]{birth} \hlc[violet!100!white]{are} unknown . In he made his appearance in Paris as a mandolin virtuoso and was highly esteemed , both as a \hlc[violet!95!white]{performer} and teacher \hlc[violet!6!white]{.} He \hlc[violet!6!white]{published} his "Method fo & 14 & 7 & 13 & -- \\
\bottomrule
\end{tabular}
\end{table*}

\section{Algorithm Pseudocode}
\label{sec:algorithm_appendix}

\begin{algorithm*}[t]
\caption{HeadRank Training and Inference}
\label{alg:attn_headrank}
\begin{algorithmic}[1]
\Require Query set $\mathcal{Q}$; BM25 top-$N$ candidates per query; pre-trained LLM $\theta_0$
\Ensure Ranking permutation $\pi$ for each query
\Statex
\Statex \textbf{Phase 1: ALPS Data Construction}
\For{each query $q \in \mathcal{Q}$}
    \State Retrieve top-100 documents via BM25 \Comment{Training: top-100}
    \State Construct preference pairs $(d_w, d_l)$ using only \emph{adjacent} relevance levels
\EndFor
\Statex
\Statex \textbf{Phase 2: Initial Core Head Selection}
\For{each head $(l,h)$ in the Transformer}
    \State Compute discriminability $S_{\text{disc}}^{(l,h)}$ and entropy gate $G_{\text{ent}}^{(l,h)}$ using the constructed data
    \State Compute selection score $\Phi(l,h) = S_{\text{disc}}^{(l,h)} \cdot G_{\text{ent}}^{(l,h)}$ \Comment{Eq.~\eqref{eq:entropy_selection}}
\EndFor
\State Select top-$K$ heads by $\Phi$; set $l_{\max} \gets \max\{l : (l,h) \in \text{top-}K\}$
\Statex
\Statex \textbf{Phase 3: HeadRank Training}
\State Freeze reference model $\theta_{\text{ref}} \gets \theta_0$
\For{each mini-batch of preference pairs}
    \State Extract attention scores $s_\theta(d)$ from core heads for $d_w, d_l$
    \State Compute $\mathcal{L}_{\text{total}} = \mathcal{L}_{\text{align}}(\Delta s_\theta) + \mathcal{L}_{\text{prox}}(\bm{\Delta}_{\text{ref}}) + \Omega(\bm{s}_\theta)$
    \State Update $\theta$ via gradient descent on $\mathcal{L}_{\text{total}}$
\EndFor
\Statex
\Statex \textbf{Phase 4: Iterative Head Recalibration}
\State Re-run \textbf{Phase 2} on updated $\theta$ to recalibrate the core head set and $l_{\max}$
\Statex
\Statex \textbf{Phase 5: Inference (Early-Exit)}
\For{each test query $q$}
    \State Run forward pass truncated at layer $l_{\max}$ \Comment{Inference: top-40}
    \State Extract attention scores from core heads; rank by $s_\theta(d)$
\EndFor
\State \Return ranking permutation $\pi$
\end{algorithmic}
\end{algorithm*}

\end{document}